\documentclass[fleqn,usenatbib]{mnras}

\usepackage{newtxtext,newtxmath}
\usepackage[T1]{fontenc}
\usepackage{graphicx}
\usepackage{amsmath}
\usepackage{booktabs}
\usepackage{xcolor}
\usepackage{orcidlink}

\title[Lensing recovery in edge-on discs]{What can strong lensing recover in edge-on disc galaxies? A mock study based on TNG50 simulations}

\author[Zheng Fang et al.]{
Zheng Fang\orcidlink{0009-0005-8336-0873}$^{1}$,
Xiaoyue Cao\orcidlink{0000-0003-4988-9296}$^{2}$\thanks{E-mail: xycao@nao.cas.cn},
Ran Li\orcidlink{0000-0003-3899-0612}$^{3}$,
Weipeng Lin\orcidlink{0000-0003-2204-2474}$^{1,4}$\thanks{E-mail: linweip5@mail.sysu.edu.cn}
\\
$^{1}$School of Physics and Astronomy, Sun Yat-sen University, DaXue Road 2, 519082, Zhuhai, People's Republic of China\\
$^{2}$Institute for Astrophysics, School of Physics, Zhengzhou University, Zhengzhou 450001, China\\
$^{3}$School of Physics and Astronomy, Beijing Normal University, Beijing 100875, China\\
$^{4}$CSST Science Center for the Guangdong-Hongkong-Macau Greater Bay Area, DaXue Road 2, 519082, Zhuhai, People's Republic of China
}

\date{}
\pubyear{\the\year{}}

\begin{document}
\label{firstpage}
\pagerange{\pageref{firstpage}--\pageref{lastpage}}
\maketitle

\begin{abstract}
Strong gravitational lensing tightly constrains the projected mass within
the Einstein aperture, but what else imaging alone can recover for edge-on
disc lenses remains untested against known truth. We construct HST/ACS-like
mock observations of 35 edge-on TNG50 disc galaxies selected to resemble
lenses from the Sloan WFC Edge-on Late-type Lens Survey (SWELLS), and fit
them with a five-stage Source--Light--and--Mass workflow in \textsc{PyAutoLens}; the
final model represents the lens mass distribution by scaling the
Multi-Gaussian Expansion (MGE) foreground-light template with a global mass-to-light ratio and adding a
spherical NFW halo.
Decomposed fits succeed for 21 of the 35 systems; in the other 14, coherent
MGE lens-light residuals prevent reliable modelling. For the successful
subset, the Einstein radius and the total aperture mass are recovered with
normalised median absolute deviation (NMAD) scatters of \ensuremath{0.3\%} and \ensuremath{0.7\%}.
The global projected mass-weighted slope has an NMAD scatter of \ensuremath{10.3\%},
smaller than that of the mass-sheet-invariant reduced deflection curvature
\(\xi\) (\ensuremath{11.4\%}) and well below that of the local slope (\ensuremath{40.7\%}), while the
stellar and dark-matter aperture masses scatter by \ensuremath{15.1\%} and \ensuremath{108.8\%}. A
35-system true-light diagnostic using the exact input light gives
component-mass scatters of \ensuremath{0.8\%} and \ensuremath{6.3\%}. Under these controlled
assumptions, imaging constrains Einstein-scale total mass and global radial
structure more reliably than local radial structure or the
stellar--dark-matter decomposition.
\end{abstract}


\section{Introduction}
\label{sec:intro}

Understanding how baryons assemble within dark-matter haloes is a central goal
of galaxy-formation theory \citep{2010PhR...495...33B}.  Disc galaxies
provide a particularly informative laboratory for this problem because their
rotationally supported structures
are shaped by cosmological gas accretion, star formation and feedback,
angular-momentum redistribution, and the response of the surrounding dark-matter halo
\citep{MoMaoWhite1998,SomervilleDave2015}.  The relative distributions of
stars and dark matter within disc galaxies therefore carry information about
how their discs, bulges, and haloes assembled and subsequently evolved.

Observationally, however, separating these mass components remains difficult.
Gas rotation curves and stellar kinematics constrain the total mass
distribution, but decomposing it into stellar and dark-matter components
requires assumptions about the stellar mass-to-light ratio, the dynamical
state of the gaseous and stellar tracers, and the dark-matter density profile
\citep{Courteau2014GalaxyMasses}. Different combinations of these components
can therefore reproduce similar kinematic observables, giving rise to the
long-standing disc--halo degeneracy
\citep{vanAlbadaSancisi1986}. This ambiguity is
particularly important
in galaxy centres, where the stellar component is often dominant and a modest
change in its normalisation can imply a much larger fractional change in the
inferred dark-matter contribution \citep{Dutton2011SWELLS}. An independent
mass probe with different geometric sensitivities is therefore valuable for
separating the two components.

Strong gravitational lensing provides such a probe when a foreground disc
galaxy deflects the light of a more distant, extended source into multiple
images or arcs.  The image configuration tightly constrains the projected
total mass enclosed by the tangential critical curve, without requiring the
lens galaxy to be in dynamical equilibrium \citep{Treu2010}.  Highly inclined
discs are especially informative because their enhanced projected surface
density and strongly flattened stellar distribution differ geometrically from
a rounder dark-matter halo; the image positions and shapes can consequently
depend on the disc mass fraction \citep{Dutton2011SWELLS}.  The Sloan WFC
Edge-on Late-type Lens Survey (SWELLS) exploited this geometry by assembling a
sample of highly inclined late-type lenses and combining strong-lensing
measurements with gas and stellar kinematics or stellar-population constraints
\citep{Treu2011SWELLS,Dutton2011SWELLS,Brewer2012SWELLS,
Barnabe2012SWELLS}.  These studies constrained the stellar and dark-matter
distributions in individual spiral lenses and disfavoured a uniformly heavy
stellar initial mass function for lower-mass systems.  In the sample-level
analyses, however, lens imaging was commonly summarised by the mass and shape
within the critical curve, with the component separation supplied by
stellar-population or kinematic information. Euclid and CSST are expected
to discover of order \(10^5\) galaxy--galaxy strong lenses over their full
surveys \citep{Collett2015,Cao2023CSSTForecast}. Euclid Q1 has already yielded
497 galaxy--galaxy strong-lens candidates, including approximately 40 edge-on
disc-lens candidates \citep{EuclidWalmsley2026,Lines2025Euclid}. If this fraction 
holds across the full survey, Euclid could detect of order \(10^4\) edge-on disc lenses. 
Comparable kinematic data will not be available for every system, motivating a clear 
assessment of what lens imaging can recover on its own.

Extended arcs contain more information than the Einstein-aperture mass alone.
Repeated source structure is mapped across a finite radial and azimuthal
region of the image plane, allowing the data to constrain spatial variations
in the lens mapping and hence aspects of the radial mass profile.  Extracting
this information nevertheless requires the unknown source morphology, the
foreground light, and the lens mass distribution to be inferred together.
Modelling images from which the lens-galaxy light has been pre-subtracted
\citep[e.g.][]{Barnabe2012SWELLS} does not propagate covariance among
foreground lens light, lensed-source light, and lens mass through the inference.
This coupling may be particularly important for edge-on discs, whose bright,
thin light distributions overlap the lensed arcs and are also used as spatial
templates for the stellar mass.

Recent developments have increased the flexibility with which this coupling can be treated in joint strong-lens analyses. 
Semi-linear inversion with adaptive pixelisation and
regularisation can reconstruct complex source morphologies
\citep{NightingaleDye2015,Nightingale2018AutoLens}.  For the foreground
galaxy, multiple MGE sets can represent radial changes in flattening,
isophotal twists, and other angular structure more flexibly than a small number
of S\'ersic components \citep{He2024MGE}.  Their intensities can be solved
alongside those of the pixelised source within a common semi-linear framework,
while automated workflows link the source-light, foreground-light, and mass
inferences across successive stages \citep{Nightingale2021PyAutoLens, 
Etherington2022NoLensLeftBehind}.  Such methods have enabled decomposed mass
models for massive elliptical lenses and flexible measurements of their total
density profiles \citep{Nightingale2019GalaxyStructure}.  Whether an imaging-only strong-lensing
analysis can recover the component masses of edge-on disc lenses reliably,
and which inferred quantities fail first when the foreground light is
imperfectly represented, remain to be established in a controlled sample with
known mass distributions.

We address this question using 35 mock edge-on disc lenses constructed from
the TNG50 simulation
\citep{Pillepich2019TNG50,Nelson2019TNG50,Nelson2019TNG}.  We select galaxies
with masses and velocity dispersions broadly comparable to SWELLS, project
them at an inclination of \(90^\circ\), and generate HST/ACS-like images. 
We fit these images with the Source--Light--and--Mass (SLaM) workflow in \textsc{PyAutoLens} \citep{Etherington2022NoLensLeftBehind, Cao2022EPLSystematics}.
In the final mass-decomposition stage, the stellar mass follows the MGE foreground-light model 
under a single spatially constant mass-to-light ratio, whereas the dark matter follows an 
NFW profile \citep{Navarro1996}. The background source is reconstructed with a pixelised model. 
Using the known input mass distributions, we assess the recovery of enclosed quantities, 
including the total, stellar, and dark-matter masses within the Einstein aperture, 
together with local radial-profile diagnostics near the Einstein radius. 
Finally, we investigate systems for which the decomposed model fails, examine the factors 
responsible for these failures, and use them to identify the practical limits 
of the adopted lens model when it is applied to real lens systems.

Section~\ref{sec:data} describes the TNG50 simulation and mock-data generation,
while Section~\ref{sec:method} summarises the strong-lens modelling methodology
adopted in this work. Section~\ref{sec:results} compares the lens-modelling
inferences with the known input values used to generate the mocks.
Section~\ref{sec:discussion} interprets variations in recovery precision and
discusses the modelling limitations and implications for observed disc lenses.
Section~\ref{sec:conclusion} summarises the main conclusions. Throughout this work, we adopt a flat \(\Lambda\)CDM cosmology consistent with
the \textit{Planck} 2018 results, with
\(H_0=67.66\,\mathrm{km\,s^{-1}\,Mpc^{-1}}\) and
\(\Omega_{\rm m,0}=0.30966\) \citep{Planck2020}.

\section{Data}
\label{sec:data}


In this section, we describe how edge-on disc galaxies with properties
broadly comparable to those of the SWELLS Grade-A lenses
\citep{Treu2011SWELLS,Brewer2012SWELLS} are selected from TNG50 and used to
generate \textit{HST}/ACS-like mock lens images.
\subsection{IllustrisTNG}
\label{subsec:tng}

IllustrisTNG is a suite of magnetohydrodynamical cosmological
simulations of galaxy formation and evolution
\citep{Pillepich2018TNGModel,Nelson2019TNG}.
The simulations employ the moving-mesh code \textsc{Arepo} \citep{Springel2010}, 
which uses a particle--mesh tree algorithm for gravity and an adaptive 
finite-volume scheme to solve the equations of ideal magnetohydrodynamics.

Each simulation evolves from $z=127$ to $z=0$
under a cosmology consistent with the \textit{Planck 2015} results
\citep{Planck2016}:
$\Omega_{\Lambda}=0.6911$, $\Omega_{\mathrm{m}}=0.3089$,
$\Omega_{\mathrm{b}}=0.0486$, $\sigma_8=0.8159$, $n_{\mathrm{s}}=0.9667$,
and $h=0.6774$.
Dark-matter haloes are identified with the Friends-of-Friends (FoF)
algorithm \citep{Davis1985} using a linking length $b=0.2$, and
gravitationally bound substructures are subsequently extracted with
\textsc{Subfind} \citep{Springel2001}. Galaxy assembly histories are
then tracked with the \textsc{Sublink} merger-tree algorithm
\citep{RodriguezGomez2015}.

The project encompasses three simulation volumes---TNG300, TNG100, and
TNG50---with box sizes of approximately 300, 100, and 50\,Mpc,
respectively.
In this work we adopt TNG50-1, the highest-resolution run, which
achieves a baryonic mass resolution of
$8.5\times10^{4}\,M_\odot$ and a dark-matter particle mass of
$4.5\times10^{5}\,M_\odot$; at $z=0$ the gravitational softening
length is $0.288\,\mathrm{kpc}$.
TNG50 captures the morphological transformation of disc galaxies from
turbulent, disordered gas clouds in the early Universe into ordered,
rotation-supported systems at low redshift. The simulation reproduces thin
discs together with bars, bulges, and spiral arms, in good agreement with
observed sizes, morphologies, and kinematics \citep{Pillepich2019TNG50}.

\subsection{Disc-galaxy selection}
\label{subsec:selection}

To construct the parent sample, we select rotationally supported TNG50
systems in a stellar-mass range that overlaps that of the SWELLS Grade-A
lenses \citep{Treu2011SWELLS,Brewer2012SWELLS}, orient them edge-on, and
apply morphological and lensing selection criteria.
All simulated galaxies are drawn from snapshot~84 of TNG50, corresponding to
a simulation redshift of \(z_{\rm snap}=0.2\). We initially require
\(10^{10.5}\le M_\star/M_\odot\le10^{11}\), where \(M_\star\) is the
total mass of all gravitationally bound stellar particles assigned by
\textsc{Subfind} to the subhalo. For each qualifying subhalo, we retrieve the
bound stellar, dark-matter, and gas particles.

We calculate the stellar angular-momentum axis from bound stellar particles
within 30~kpc, using positions and velocities relative to the catalogued subhalo
position and bulk velocity. This axis defines the azimuthal velocity used to
quantify rotational support. Following the motivation of
\citet{Sales2010Feedback}, we identify disc-dominated systems with the
mass-weighted statistic adopted by \citet{Zhao2020Bars},
\begin{equation}
\kappa_{\rm rot}=
\frac{1}{M_{\star,30}}
\sum_{r_i<30\,{\rm kpc}}
m_i\frac{v_{\phi,i}^2}{v_i^2},
\label{eq:kappa_rot}
\end{equation}
where \(M_{\star,30}\) is the total stellar mass within this aperture, and
\(v_{\phi,i}\) and \(v_i\) are the azimuthal and total speeds of particle
\(i\), respectively. We require \(\kappa_{\rm rot}\ge0.45\). This threshold
is slightly less restrictive than the value of 0.5 adopted by
\citet{Zhao2020Bars} and permits the target sample to include rotationally
supported systems with substantial central bulges.

After the kinematic selection, we centre each galaxy on an iteratively
determined stellar-mass centre and align it with the principal axes of the
bound stellar mass distribution. We determine these axes using a
mass-weighted extension of the iterative reduced inertia tensor method of
\citet{Allgood2006Shapes},
\begin{equation}
I_{ij}=\sum_k \frac{m_k x_i^{(k)}x_j^{(k)}}{r_k^2},
\label{eq:inertia}
\end{equation}
where the sum is evaluated over bound stellar particles inside an
iteratively updated ellipsoidal aperture. Its major-axis radius is fixed to
\(R_b=5R_{1/2}\), where \(R_{1/2}\) is the three-dimensional stellar
half-mass radius. Here \(m_k\) and \(x_i^{(k)}\) are the mass and \(i\)th
coordinate of particle \(k\), and \(r_k\) is its ellipsoidal distance from
the stellar centre. The eigenvectors of \(I_{ij}\) define the principal-axis
directions, and the iteration terminates when the changes in both \(b/a\)
and \(c/a\) are below 0.01. In the resulting frame, the XY projection is
face-on and the XZ and YZ projections are edge-on.

In this coordinate frame, we use an adaptive kernel-smoothing scheme based on smoothed particle hydrodynamics (SPH) to project the bound stellar, dark-matter, and gas mass distributions along the three principal axes. We adopt a spherical \(\mathrm{M}_4\) cubic-spline kernel \citep{Monaghan1992SPH} and integrate it along the line of sight to evaluate the projected surface-density contribution of each particle or gas cell \citep{Price2007SPLASH}. The resulting component maps are then summed to obtain the total surface-density maps.
The stellar luminosity is projected onto the same grid and in the same
orientation. We visually inspect
these edge-on luminosity maps and exclude systems with irregular
morphologies or poor symmetry. This step retains
regular disc structures suitable for the controlled photometric-modelling
experiment.

\begin{figure*}
    \centering
    \includegraphics[width=0.9\textwidth]{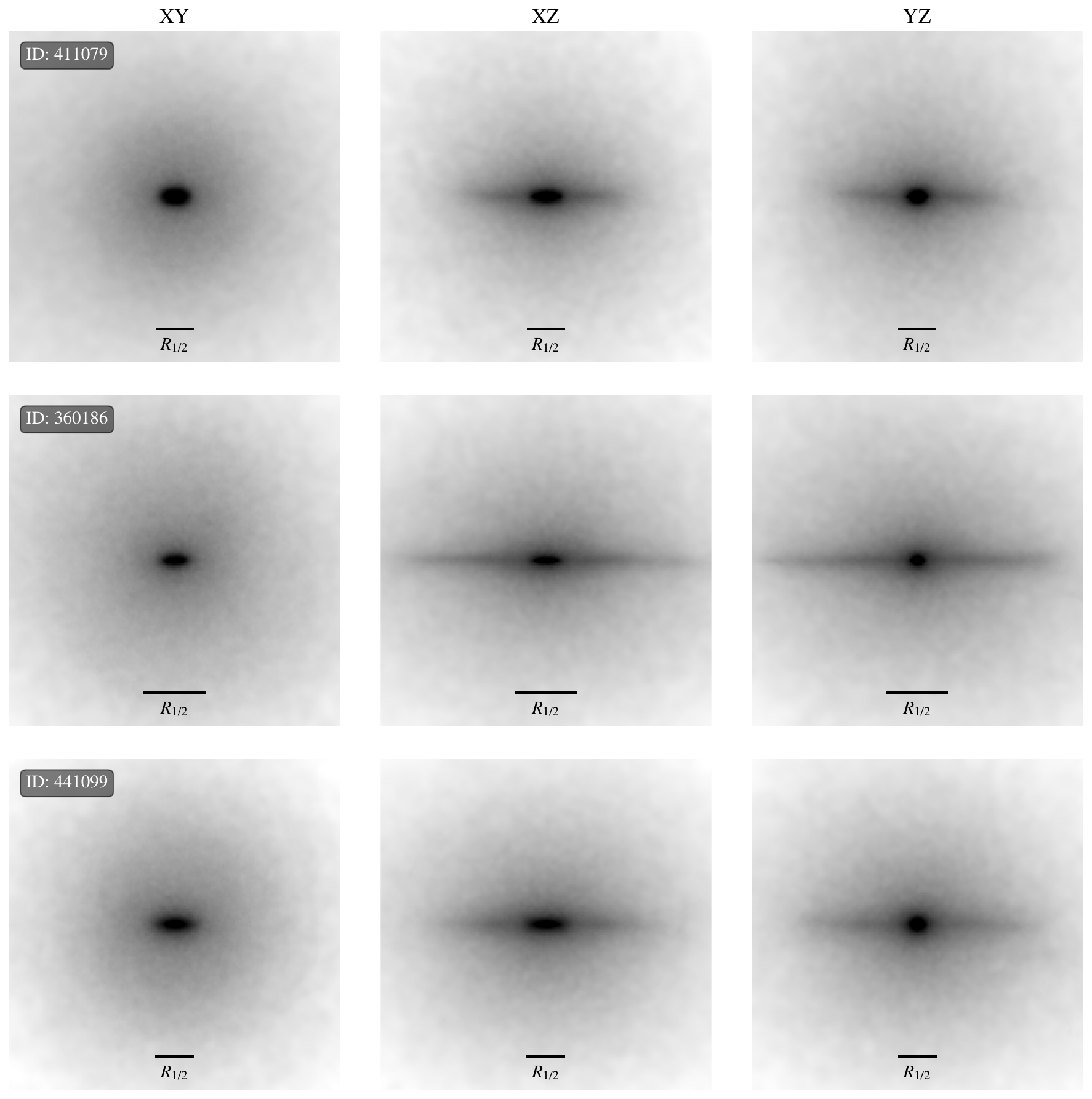}
    \caption{
    Three disc galaxies from the final selected TNG50 sample. The galaxies are drawn from
    TNG50 snapshot~84 (\(z_{\rm snap}=0.2\)), and the panels show their
    particle-based total mass distributions, including the stellar,
    dark-matter, and gas components, projected along the three principal
    axes. Galaxy IDs are indicated in the upper-left corners. Each image
    covers a region of 30~kpc on a side, and the black bar denotes the stellar
    half-mass radius \(R_{1/2}\). \textit{Left:} face-on;
    \textit{middle/right:} edge-on.
    }
    \label{fig:mass_projection}
\end{figure*}

For the strong-lensing mock generation, we place each simulated TNG50 galaxy
at a common lens redshift of \(z_{\rm l}=0.15\) and adopt
\(z_{\rm s}=0.65\) for the background source. These redshifts are
representative of the lens and source redshifts in the SWELLS sample
\citep{Treu2011SWELLS,Brewer2012SWELLS}. From the combined particle-based
mass map, we derive the convergence and shear fields and identify the
tangential critical curve. We define the effective Einstein radius as
\(\theta_{\rm E}=(A_{\rm crit}/\pi)^{1/2}\), where \(A_{\rm crit}\) is the
angular area enclosed by this curve
\citep{Puchwein2009EinsteinRadius}. We retain systems with \(\theta_{\rm E}\geq0.5\arcsec\), because smaller
Einstein radii place the multiple images closer to the lens centre, where
blending with the foreground lens light makes them more difficult to identify
and model reliably at \textit{HST} resolution.

\begin{figure*}
  \centering
  \includegraphics[width=0.95\textwidth]{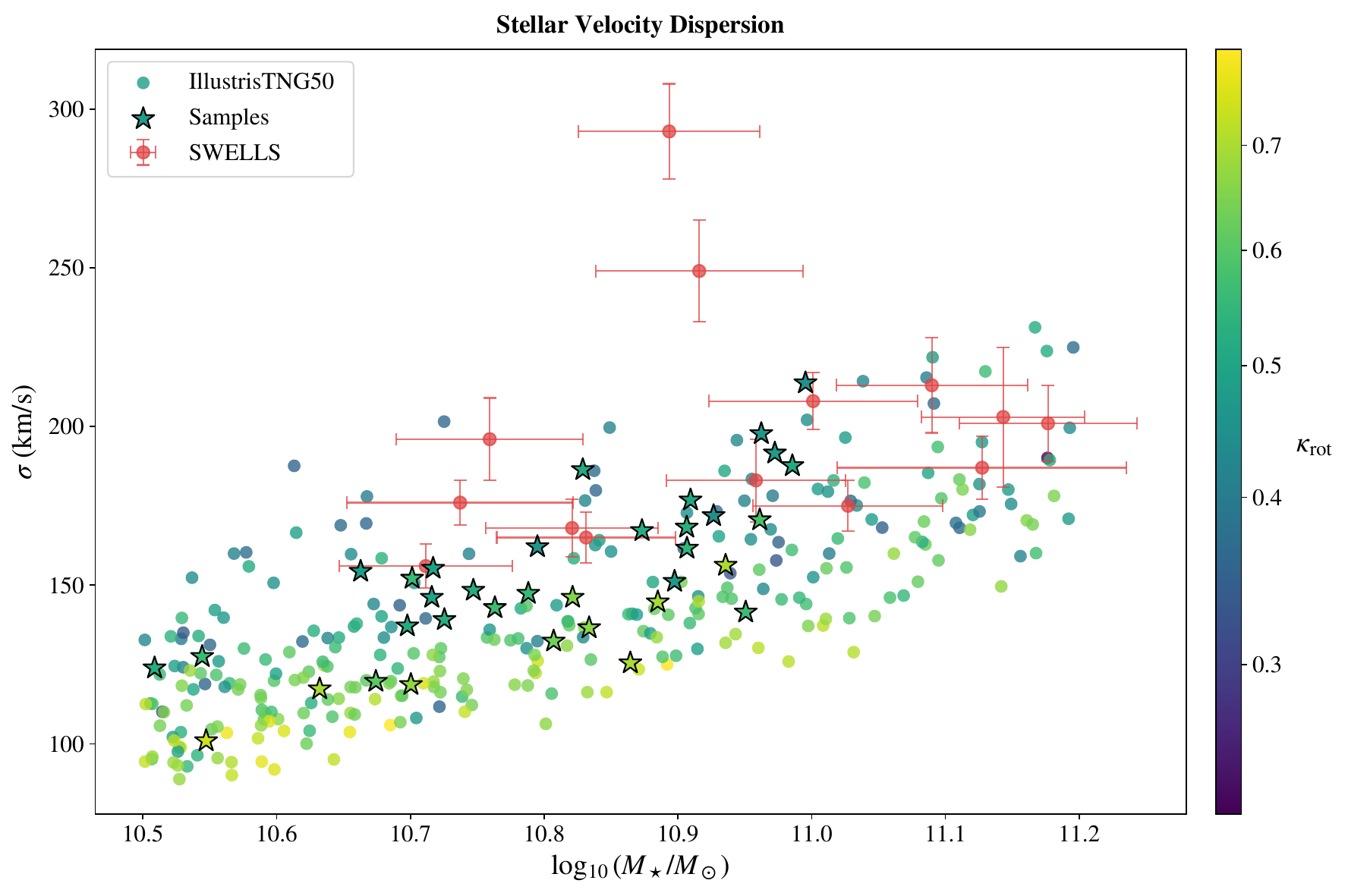}
  \caption{
  Stellar velocity dispersion versus stellar mass for the 35 selected TNG50
  disc galaxies (stars), the parent TNG50 population (circles), and the
  SWELLS Grade-A lenses (red points). The colour bar encodes
  \(\kappa_{\rm rot}\), the dimensionless rotational-support statistic
  defined in equation~\ref{eq:kappa_rot}; larger values indicate stronger
  rotational support. The selected galaxies satisfy
  \(\kappa_{\rm rot}\geq0.45\) and
  \(10^{10.5}\le M_\star/M_\odot\le10^{11}\).
  }
  \label{fig:dispersion}
\end{figure*}

The final parent sample contains 35 disc galaxies. As a qualitative check,
Fig.~\ref{fig:dispersion} compares their stellar masses and velocity
dispersions with those of the SWELLS Grade-A lenses
\citep{Treu2011SWELLS,Brewer2012SWELLS}.

While the selected TNG50 discs sample the lower-\(M_\star\) and lower-\(\sigma\) end of the SWELLS distribution, they span a comparable regime in the \(M_\star\)--\(\sigma\) plane, providing an observational reference for our mock sample.

\subsection{Mock lens image generation}
\label{subsec:mock}

For each selected galaxy, we construct a mock lens image at a lens
redshift of \(z_{\rm l}=0.15\). Using the edge-on orientation defined
above, we project the TNG50 stellar particles
along the principal X axis onto the YZ plane with the  adaptive kernel-smoothing scheme,
producing stellar-mass and luminosity maps at the \textit{HST}/ACS pixel scale of
\(0.05\arcsec\). The intrinsic SDSS \(i\)-band particle luminosities are
converted first to apparent magnitudes using the luminosity distance and
then to count rates using an ACS/F814W AB zero point of 25.948
\citep{STScIACSZeropoints}. The resulting foreground-light map is
expressed in \(\mathrm{counts\,s^{-1}\,pixel^{-1}}\).

The lens mass model combines a stellar component with an idealised
dark-matter halo. For each system, the projected light distribution is
scaled by the aperture-averaged stellar mass-to-light ratio within the
 input Einstein radius to produce a stellar convergence map. We derive
the corresponding deflection field by convolving this map with the
lensing deflection kernel using an FFT, and supply the result to
\textsc{PyAutoLens} as a custom stellar mass component. The finite
resolution of the convergence map can artificially reduce the central
deflection and produce a spurious central image. We suppress this
numerical artefact by assigning a large deflection within
\(0.2\arcsec\), thereby mapping those image-plane pixels far from the
source. For the dark matter, we replace the simulated particle
distribution with a spherical Navarro--Frenk--White (NFW) profile to
avoid an artificial central core caused by finite numerical resolution
and gravitational softening
\citep{Navarro1996,Mukherjee2018SEAGLE,Ding2021TDLMC}. The halo mass
\(M_{200}\) and radius \(r_{200}\), defined relative to 200 times the
critical density, are measured directly from the simulation. Its scale
radius is set by the median mass--concentration relation of
\citet{Ludlow2016}, with no intrinsic scatter, and the halo is implemented
as a \texttt{NFWMCRLudlowSph} profile in \textsc{PyAutoLens}.

We place a single S\'ersic source \citep{Sersic1968} at
\(z_{\rm s}=0.65\), with \(r_{\rm e}=0.15\arcsec\), \(n=1.0\),
\(q=0.7\), and position angle \(\phi_{\rm s}=0^\circ\). Its centre is
offset from the lens centre by a Gaussian draw with a standard deviation
of \(0.1\arcsec\). We ray-trace this source through the
combined stellar and NFW deflection field and add the foreground-galaxy
light to form the noiseless mock image.

To reproduce an \textit{HST}/ACS F814W-like observation, we convolve the
noiseless image with a Gaussian PSF of standard deviation
\(0.05\arcsec\), adopt a sky background of
\(0.1\,\mathrm{counts\,s^{-1}\,pixel^{-1}}\), and set the exposure time to
\(2400\,\mathrm{s}\). We then add the sky background and Poisson noise
from the target source. The source brightness \(I_{\rm e}\) is adjusted
so that each mock has a peak signal-to-noise ratio of
\(\mathrm{S/N}\simeq50\). These observational settings follow
\citet{Cao2022EPLSystematics}.

\section{Method}
\label{sec:method}

We model the simulated images with \textsc{PyAutoLens}
\citep{Nightingale2018AutoLens,Nightingale2021PyAutoLens}. Galaxy light
profiles are defined with \textsc{PyAutoGalaxy}
\citep{Nightingale2023PyAutoGalaxy}, whereas Bayesian parameter inference
is performed with \textsc{PyAutoFit} \citep{Nightingale2021PyAutoFit},
using either the \texttt{nautilus} \citep{Lange2023Nautilus} or
\texttt{dynesty} \citep{Speagle2020Dynesty} nested sampler.

\subsection{SLaM modelling workflow}
\label{subsec:slam}

The joint inference of the foreground lens light, lensed source emission,
and lens mass from the same image-plane data introduces partial
degeneracies and a complex parameter space. We manage this complexity
using five linked searches in the Source--Light--and--Mass (SLaM)
workflow, with constraints from earlier searches informing those that
follow
\citep{Etherington2022NoLensLeftBehind,He2024MGE}. The workflow
progressively increases the flexibility of the source, foreground-light,
and mass models, while using constraints from earlier searches to
initialise or restrict the models fitted subsequently.

The Source Light Profile search (SP) first jointly fits parametric models for the
foreground lens light, total lens mass, and background source, thereby
establishing the initial lens mapping and lens--source decomposition.
Source Pix 1 (SPix1) replaces the parametric source with an initial
pixelised reconstruction and refines the total-mass model using priors
from SP. Source Pix 2 (SPix2) then uses this initial reconstruction to
adapt the source pixelisation and regularisation, while holding the lens
mass fixed to the SPix1 solution. The Light Profile search (L) subsequently
refits the foreground-light model using the SPix1 lens mapping and the
source-model configuration established in SPix2. Finally, the Mass
Light Dark search (M) replaces the initial total-mass model with separate
stellar and dark-matter components.

The specific foreground-light, source, and lens-mass parameterisations
are described in Sections~\ref{subsec:light}--\ref{subsec:mass_model}
and summarised in Table~\ref{tab:slam}.

A \texttt{PositionsLHPenalty} is applied in searches where the lens
mass parameters are varied. For each system, two to four positions are
manually selected from bright regions along the lensed arcs. Mass models
that fail to map these positions to a common source-plane location within
the specified threshold are penalised, helping to suppress unphysical
demagnified solutions \citep{Nightingale2018AutoLens}. Bayesian evidence
is used to evaluate the non-linear fits and source regularisation,
accounting for both goodness of fit and effective model complexity.

\subsection{Foreground lens-light model}
\label{subsec:light}

The foreground light of the lens galaxy is modelled with a
Multi-Gaussian Expansion (MGE)
\citep{Cappellari2002MGE,He2024MGE}. Following the set-based
parameterisation of \citet{He2024MGE}, the surface-brightness
distribution is written as
\begin{equation}
I_{\rm lens}^{\rm MGE}(x,y)
=
\sum_{i=1}^{N_{\rm b}}
\sum_{j=1}^{N_{\rm g}^{i}}
I_{i,j}
\exp\left[-\frac{R_i^2(x,y)}{2\sigma_{i,j}^2}\right],
\label{eq:mge}
\end{equation}
where \(I_{i,j}\) and \(\sigma_{i,j}\) are the intensity and width of the
\(j\)th Gaussian in set \(i\). All Gaussians within a set share the
same elliptical radius,
\begin{equation}
\begin{aligned}
R_i^2(x,y)
&={x_i'}^2+\left(\frac{y_i'}{q_i}\right)^2,\\
x_i'
&=(x-x_0)\cos\theta_i+(y-y_0)\sin\theta_i,\\
y_i'
&=-(x-x_0)\sin\theta_i+(y-y_0)\cos\theta_i,
\end{aligned}
\label{eq:mge_radius}
\end{equation}
where \(q_i\) and \(\theta_i\) are the axis ratio and position angle
of set \(i\), respectively. To facilitate sampling, these are
reparameterised in terms of the two ellipticity components
\begin{equation}
e_{1,i}=\frac{1-q_i}{1+q_i}\sin(2\theta_i),
\qquad
e_{2,i}=\frac{1-q_i}{1+q_i}\cos(2\theta_i).
\end{equation}
All sets share the centroid \((x_0,y_0)\), while their ellipticity
components are fitted independently.

We model the lens-galaxy light using two MGE sets for the bulge and two
for the disc. Each set contains 15 Gaussians that share a common
ellipticity, while the ellipticity components are fitted independently
across sets. The Gaussian widths are fixed on a logarithmic grid
extending to the mask radius, and, for a given centroid and set
ellipticities, their intensities are determined by linear inversion.

The lens MGE is initially fitted together with the parametric source and
SIE in the Source Light Profile search. Its non-linear geometry is fixed during the
two Source Pix searches while the source and lens mass are refined. The
four-set MGE is then refitted in the Light Profile search using the SPix1 lens
mapping and the pixelisation and regularisation configurations from
SPix2. The resulting light model also serves as the stellar-mass template
in the final Mass Light Dark search.  

\subsection{Source reconstruction}
\label{subsec:source}

The source model progressively increases in flexibility across the
first three SLaM searches. A low-dimensional parametric source provides
a stable initial estimate of the lens mapping; a first pixelised search
then establishes the source morphology; and a second pixelised search
uses that information to adapt both the spatial resolution and
regularisation of the reconstruction.

In the Source Light Profile search, the background source is represented by one
set of ten Gaussians whose widths are logarithmically spaced from
\(10^{-3}\arcsec\) to \(1\arcsec\). The Gaussians share a common
centroid and ellipticity, while their intensities are determined by
linear inversion. Fitted jointly with the initial lens-mass model, this
restricted source representation provides an estimate of the lens
geometry and source location for the subsequent pixelised searches; it
is not intended as the final reconstruction of the source morphology.

The pixelised source is reconstructed using regularised semi-linear
inversion
\citep{WarrenDye2003,Suyu2006Bayesian,NightingaleDye2015,
Nightingale2018AutoLens}. For fixed non-linear lens parameters, the lens
mapping and PSF convolution define a linear relation between the
source-pixel brightnesses and the image-plane data. The source
brightnesses are therefore solved simultaneously by linear inversion
rather than sampled individually by the non-linear sampler. The source
pixels lie on an irregular Delaunay mesh whose vertices are obtained by
ray-tracing image-plane sampling points to the source plane. The local
source-plane vertex density consequently sets the effective resolution
of the reconstruction.

The first Source Pix search uses a \(30\times30\) Overlay image mesh to
define the source pixelisation. Uniformly distributed image-plane
sampling points are ray-traced through the lens model and used as the
vertices of a Delaunay triangulation in the source plane. As this
sampling does not depend on a previous source reconstruction, the mesh
is not adapted to the source brightness. This search provides an initial
pixelised reconstruction of the source while refining the SIE and
external shear using priors from the Source Light Profile result.

The second Source Pix search uses the source morphology inferred in the
first search to construct an adaptive reconstruction at two levels.
First, a 1000-point Hilbert image mesh adapts the sampling to the
reconstructed source emission, allocating greater source-plane
resolution to the regions that contribute most strongly to the lensed
image. Secondly, neighbouring source pixels are coupled using
\texttt{AdaptiveBrightnessSplit} regularisation, which applies weaker
smoothing to bright source structures and stronger smoothing to faint
regions. This suppresses noise fitting while retaining resolved source
structure. The regularisation strength is constrained within the
Bayesian-evidence framework \citep{Suyu2006Bayesian}. The second search
therefore adapts both the allocation of source pixels and the smoothing
applied across them.

The pixelisation and regularisation configurations established in the
second Source Pix search define the source model used in the subsequent
Light Profile and Mass Light Dark stages. The source-pixel brightnesses are
not retained as a fixed reconstruction, but are re-solved by
semi-linear inversion within each subsequent fit.

\subsection{Lens mass models}
\label{subsec:mass_model}

\subsubsection{Initial total-mass model}

The first four searches use a Singular Isothermal Ellipsoid (SIE)
\citep{Kormann1994} with external shear to describe the total lens mass.
In the Source Light Profile search, the SIE centre is initially fixed to the image
centre, and the model is fitted jointly with the parametric
source to establish the initial lens geometry. Source Pix 1 retains the
same mass parameterisation but refines its parameters using priors from
the Source Light Profile result while replacing the parametric source with the
initial pixelised reconstruction. The resulting SIE and shear are held
fixed in Source Pix 2, so that the adaptive source reconstruction is
constructed for a fixed lens mapping, and in Light Profile, where the
foreground light is refitted.

\subsubsection{Final stellar and dark-matter model}

After the foreground light has been refined in the Light Profile search, the
Mass Light Dark search replaces the SIE with a decomposed stellar and
dark-matter model. The fitted MGE surface-brightness distribution is
converted into the stellar mass component using a single global
mass-to-light-ratio parameter in \textsc{PyAutoLens}. Its centroid, axis
ratios, and position angles are inherited from the Light Profile result,
while the global mass-to-light ratio is fitted.

The dark-matter component is described by a spherical
\texttt{NFWMCRLudlowSph} profile, which assigns the halo concentration
according to the mass--concentration relation of \citet{Ludlow2016}.
Its only free parameter is \(M_{200}\), and its centre is tied to the
common centroid of the stellar MGE. The total convergence of the
decomposed model is
\begin{equation}
\kappa_{\rm tot}(x,y)
=
\kappa_\star(x,y)+\kappa_{\rm NFW}(x,y),
\end{equation}
with the external shear retained as a separate contribution to the
lensing deflection.

\begin{table*}
  \centering
  \caption{Summary of the five-stage SLaM modelling workflow, showing the
foreground lens-light, lens-mass, and source models adopted in each
search together with the corresponding nested sampler. Components
listed as fixed are inherited from the indicated earlier search.}
  \label{tab:slam}
  \small
  \renewcommand{\arraystretch}{1.12}
  \begin{tabular*}{\textwidth}{l@{\extracolsep{\fill}}llll}
    \toprule
    Stage & Lens light & Lens mass & Source light & Sampler \\
    \midrule
    SP
      & Four-set MGE
      & SIE + shear
      & One-set MGE
      & \texttt{nautilus} \\
    SPix1
      & Geometry fixed from SP
      & SIE + shear, SP priors
      & Overlay--Delaunay
      & \texttt{nautilus} \\
    SPix2
      & Geometry fixed from SP
      & Fixed from SPix1
      & Adaptive Hilbert--Delaunay
      & \texttt{dynesty} \\
    L
      & Four-set MGE, refitted
      & Fixed from SPix1
      & SPix2 pixelisation + regularisation
      & \texttt{nautilus} \\
    M
      & MGE from L
      & MGE \(M/L\) + NFW + shear
      & SPix2 pixelisation + regularisation
      & \texttt{nautilus} \\
    \bottomrule
  \end{tabular*}
\end{table*}

\subsection{True-light diagnostic}
\label{subsec:true_light_control}

The fitted MGE describes both the foreground lens light and the spatial
template of the stellar mass. Residuals in this light model may therefore
propagate into the inferred stellar and dark-matter parameters. We
perform a computationally inexpensive true-light diagnostic of mass
recovery when the foreground-light distribution is known exactly.

We subtract the exact input foreground light from each mock image and
construct a unit-normalisation stellar deflection field from the same
light map. This field is scaled by a free global mass-to-light ratio,
while the dark-matter component remains a spherical NFW halo with free
\(M_{200}\). For simplicity and computational efficiency, the remaining
lensed-source emission is fitted with a parametric S\'ersic model from
the same functional family used to generate the mocks.

This diagnostic is not a controlled isolation of the MGE fitting
residuals, because it changes both the foreground-light treatment and
the source model relative to the main analysis. Instead, it describes
the mass recovery attainable under a simplified true-light
configuration. Isolating the effect of the MGE residuals would require
repeating the analysis with the same pixelised source model while
changing only the foreground-light and stellar-mass templates.

\begin{figure*}
    \centering
    \includegraphics[width=0.98\textwidth]{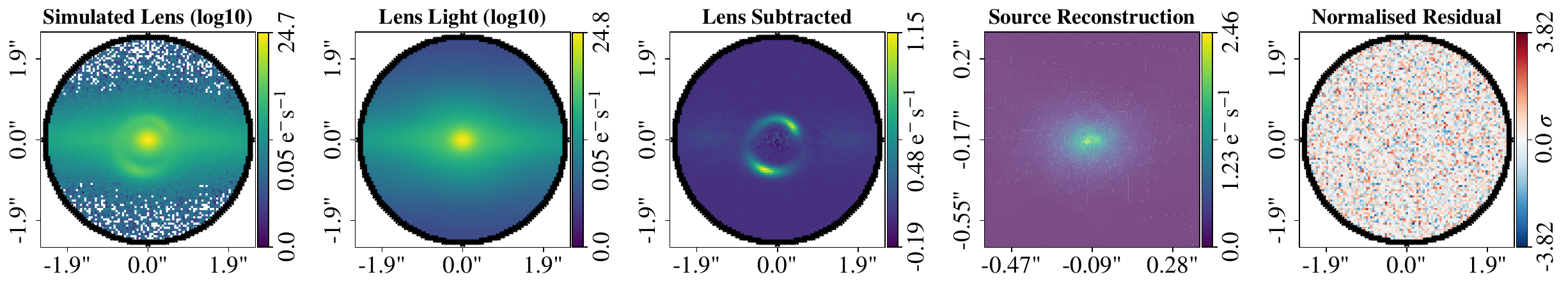}
    \caption{
    Example of the final SLaM fit for one mock lens. From left to
    right, the panels show the simulated image, the fitted lens light,
    the lens-light-subtracted image, the reconstructed source, and the
    normalised image-plane residuals. The fit is evaluated at the
    posterior median. 
    }
    \label{fig:slam_example}
\end{figure*}

\subsection{Posterior summaries and measured quantities}
\label{subsec:metrics}

All quantitative recovery statistics are derived from the relevant
\texttt{nautilus} posterior, not from a single best-fit or
maximum-likelihood point estimate. For both the main Mass Light Dark
stage and the true-light diagnostic, we converted the weighted posterior
samples to equally weighted draws using
\texttt{dynesty.utils.resample\_equal} \citep{Speagle2020Dynesty}.
For each posterior draw in the main SLaM analysis, we reconstructed the
full lens mass model and computed its effective Einstein radius from the
area enclosed by the tangential critical curve. The total, stellar, and
dark-matter masses were integrated within this draw-specific Einstein
aperture. A circularised total-convergence profile was then constructed
for the same draw, from which we evaluated the local convergence
\(\kappa(\theta_{\rm E})\), deflection curvature
\(\alpha''(\theta_{\rm E})\), reduced deflection curvature
\(\xi(\theta_{\rm E})\), global projected mass-weighted slope
\(\gamma(<\theta_{\rm E})\), and local projected mass-weighted slope
\(\gamma(\theta_{\rm E})\) using the same draw-specific radius.

For each derived quantity, the point estimate reported for a system is
the median of its posterior draws, and the associated uncertainty
interval spans the 16th to 84th percentiles. These per-system summaries
are shown as points and error bars in the recovery figures. For a
recovered quantity \(q\), we define the fractional residual as
\begin{equation}
\Delta_q=\frac{q_{\rm fit}-q_{\rm input}}{q_{\rm input}}.
\label{eq:fractional_residual}
\end{equation}
We describe the centre of the cross-system residual distribution by the
median fractional bias
\begin{equation}
b_q=\mathrm{median}(\Delta_q),
\label{eq:median_bias}
\end{equation}
where positive and negative values denote systematic overestimation and
underestimation, respectively. We quantify the scatter about this centre
using the normalised median absolute deviation (NMAD),
\begin{equation}
\sigma_{\rm NMAD}
=
1.48\,\mathrm{median}
\left(
\left|
\Delta_q-b_q
\right|
\right).
\end{equation}
The factor of 1.48 makes \(\sigma_{\rm NMAD}\) approximately equal to the
standard deviation for a Gaussian distribution, while retaining the lower
sensitivity of the median absolute deviation to outliers. Thus, \(b_q\)
measures a common directional offset, whereas \(\sigma_{\rm NMAD}\) measures
the robust system-to-system scatter around that offset. We report both
quantities as percentages throughout.

\section{Results}
\label{sec:results}

We quantify recovery from the distribution of signed fractional residuals
\(\Delta_q\) defined in equation~\eqref{eq:fractional_residual}, with each
system represented by the posterior median of its recovered quantity. Positive
and negative residuals indicate overestimation and underestimation,
respectively. All quoted \(b_q\) and \(\sigma_{\rm NMAD}\) values are
cross-system summaries reported as percentages.
Fig.~\ref{fig:slam_example} illustrates the final
SLaM fit for one mock lens.

Of the 35 mock systems, 21 yield successful decomposed fits and form the
sample used for the quantitative recovery analysis below. The reported
recovery statistics are therefore conditional on this 21-system subset.
The remaining 14 exhibit visibly inadequate foreground lens-light fits,
with coherent residuals along the disc plane that propagate through the
SLaM workflow and prevent successful recovery of the separate stellar and
dark-matter components. We discuss these systems and the limitations of
the adopted MGE lens-light model in
Section~\ref{subsec:light_discussion}.

\subsection{Einstein-aperture mass recovery}
\label{subsec:mass_recovery}

We first examine the recovery of the projected total, stellar, and
dark-matter masses---\(M_{\rm tot}(<\theta_{\rm E})\),
\(M_\star(<\theta_{\rm E})\), and
\(M_{\rm DM}(<\theta_{\rm E})\), respectively---within the effective
Einstein aperture. For each system, the  fitted and input masses are
integrated within their respective Einstein radii.

\begin{figure*}
    \centering
    \includegraphics[width=0.95\textwidth]{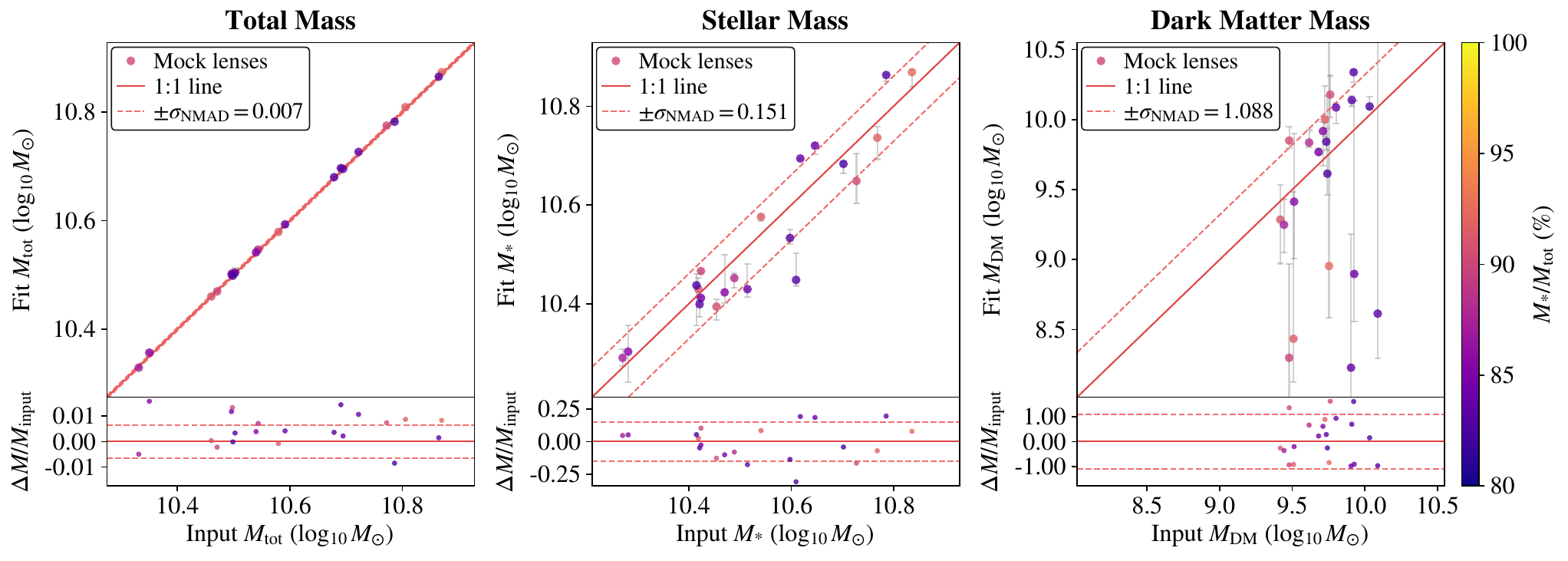}
    \caption{
    Recovered versus  input projected masses inside the Einstein aperture for
    the total, stellar, and dark-matter components in the 21 systems.
    For each system, the masses are evaluated for every posterior draw;
    points show the posterior medians and vertical bars span the 16th to
    84th percentiles.
    Points in all three panels are colour-coded by the input stellar fraction
    \(M_\star/M_{\rm tot}\) within the aperture.
    Solid lines mark the one-to-one relation; bottom subpanels show the
    signed fractional residuals used to calculate \(b_q\) and
    \(\sigma_{\rm NMAD}\). Dashed lines indicate the zero-centred
    \(\pm\sigma_{\rm NMAD}\) bounds.
    }
    \label{fig:mass_recovery}
\end{figure*}

Fig.~\ref{fig:mass_recovery} compares the recovered and  input masses for
the 21 systems with successful decomposed fits. Across these systems, the
total aperture-mass residual distribution has a median bias of \(\ensuremath{+0.39\%}\)
and \(\sigma_{\rm NMAD}=\ensuremath{0.7\%}\). The individual residuals remain
within \(\pm2\)\%. The  input total masses span approximately
\(10^{10.3}\)--\(\ensuremath{10^{10.9}}\,{\rm M}_\odot\).

Across the same systems, the stellar aperture-mass residual distribution has
a median bias of \(\ensuremath{-2.7\%}\) and \(\sigma_{\rm NMAD}=\ensuremath{15.1\%}\).
The individual residuals remain within \(\pm\ensuremath{31}\)\%. Most systems are strongly
stellar-dominated within the Einstein aperture, with
\(M_\star/M_{\rm tot}>0.8\).

The dark-matter aperture mass is recovered much less precisely. Its residual
distribution has a median bias of \(\ensuremath{+15.0\%}\) and
\(\sigma_{\rm NMAD}=\ensuremath{108.8\%}\). Both component-mass biases are smaller than
their corresponding \(\sigma_{\rm NMAD}\) values. Because the total aperture
mass is tightly anchored, a negative
shift in the dominant stellar component is compensated by a positive shift in
the subdominant dark-matter component. Expressing this shift relative to the
small dark-matter mass amplifies its fractional residual. The decomposition is
therefore limited primarily by system-to-system scatter rather than by a
common bias, although the component residuals also show a directional offset.

\begin{table}
  \centering
  \caption{Median fractional bias \(b_q\) and normalised median absolute
  deviation \(\sigma_{\rm NMAD}\) for the recovered quantities in the 21
  systems with successful decomposed fits. Positive and negative \(b_q\)
  values denote overestimation and underestimation, respectively. Both
  statistics are reported as percentages. Masses are integrated within the
  corresponding effective Einstein aperture. The local slope is evaluated
  over \([0.95\theta_{\rm E},1.05\theta_{\rm E}]\).}
  \label{tab:summary_stats}
  \small
  \renewcommand{\arraystretch}{1.12}
  \begin{tabular*}{\columnwidth}{l@{\extracolsep{\fill}}cc}
    \toprule
    Quantity & \(b_q\) (\%) & \(\sigma_{\rm NMAD}\) (\%) \\
    \midrule
    \(M_{\rm tot}(<\theta_{\rm E})\) & \(\ensuremath{+0.39}\) & \(\ensuremath{0.7}\) \\
    \(M_\star(<\theta_{\rm E})\) & \(\ensuremath{-2.7}\) & \(\ensuremath{15.1}\) \\
    \(M_{\rm DM}(<\theta_{\rm E})\) & \(\ensuremath{+15.0}\) & \(\ensuremath{108.8}\) \\
    \(\theta_{\rm E}\) & \(\ensuremath{+0.01}\) & \(\ensuremath{0.3}\) \\
    \(\gamma(<\theta_{\rm E})\) & \(\ensuremath{+0.4}\) & \(\ensuremath{10.3}\) \\
    \(\xi(\theta_{\rm E})\) & \(\ensuremath{+5.4}\) & \(\ensuremath{11.4}\) \\
    \(\alpha''(\theta_{\rm E})\) & \(\ensuremath{+4.5}\) & \(\ensuremath{21.0}\) \\
    \(\kappa(\theta_{\rm E})\) & \(\ensuremath{+2.4}\) & \(\ensuremath{34.0}\) \\
    \(\gamma(\theta_{\rm E})\) & \(\ensuremath{+1.0}\) & \(\ensuremath{40.7}\) \\
    \bottomrule
  \end{tabular*}
\end{table}

\subsection{Einstein-scale radial structure}
\label{subsec:radial_structure}
In addition to the total mass enclosed by the Einstein
aperture, the extended lensed image contains information about the radial mass profile. 
We therefore examine the recovery of \(\theta_{\rm E}\) and five
enclosed or local diagnostics derived from the circularised radial profile.

Following the convention of \citet{Kochanek2019}, we define the
dimensionless reduced deflection curvature \(\xi\) as
\begin{equation}
\xi(\theta_{\rm E}) =
\frac{\theta_{\rm E}\alpha''(\theta_{\rm E})}
{1-\kappa(\theta_{\rm E})},
\label{eq:xi_def}
\end{equation}
where \(\alpha''\) is the second radial derivative of the circularised
deflection profile and \(\kappa(\theta_{\rm E})\) is the local convergence.
This quantity measures the normalised radial curvature of the deflection
profile at the Einstein radius and is invariant under a mass-sheet
transformation \citep{SchneiderSluse2013}. When angular radii are expressed in \(\mathrm{arcsec}\),
\(\alpha''\) has units of \(\mathrm{arcsec}^{-1}\).

The global projected mass-weighted slope is the two-dimensional analogue of
the construction of \citet{DuttonTreu2014}: the mean of the local projected
slope weighted by the projected mass in each annulus from the centre to the
Einstein radius. Writing the local projected slope as
\(\gamma_\kappa(\theta)=-d\ln\kappa_{\rm tot}/d\ln\theta\), we define
\begin{align}
\gamma(<\theta_{\rm E})
&\equiv
\frac{1}{M_{\rm 2D}(<\theta_{\rm E})}
\int_0^{\theta_{\rm E}}
\gamma_\kappa(\theta)\,dM_{\rm 2D}(\theta)
\nonumber\\
&=2-\frac{\theta_{\rm E}^{2}\kappa_{\rm tot}(\theta_{\rm E})}
{\int_0^{\theta_{\rm E}}\kappa_{\rm tot}(\theta)\,\theta\,d\theta}
\nonumber\\
&=2-\left.
\frac{d\ln M_{\rm 2D}(<\theta)}{d\ln\theta}
\right|_{\theta=\theta_{\rm E}},
\label{eq:gamma_enclosed}
\end{align}
where \(M_{\rm 2D}(<\theta)=2\pi\Sigma_{\rm crit}
\int_0^\theta\kappa_{\rm tot}(x)\,x\,dx\) is the projected mass enclosed
within \(\theta\). The local projected mass-weighted slope near the Einstein
radius, \(\gamma(\theta_{\rm E})\), is instead estimated as the
projected-mass-weighted mean of \(\gamma_\kappa(\theta)\) over
\([0.95\theta_{\rm E},1.05\theta_{\rm E}]\), providing a finite-width
estimate of the local slope at \(\theta_{\rm E}\). The radial
diagnostics are evaluated from azimuthally averaged profiles at
the corresponding Einstein radius, defined from the area enclosed
by the tangential critical curve.

\begin{figure*}
    \centering
    \includegraphics[width=0.95\textwidth]{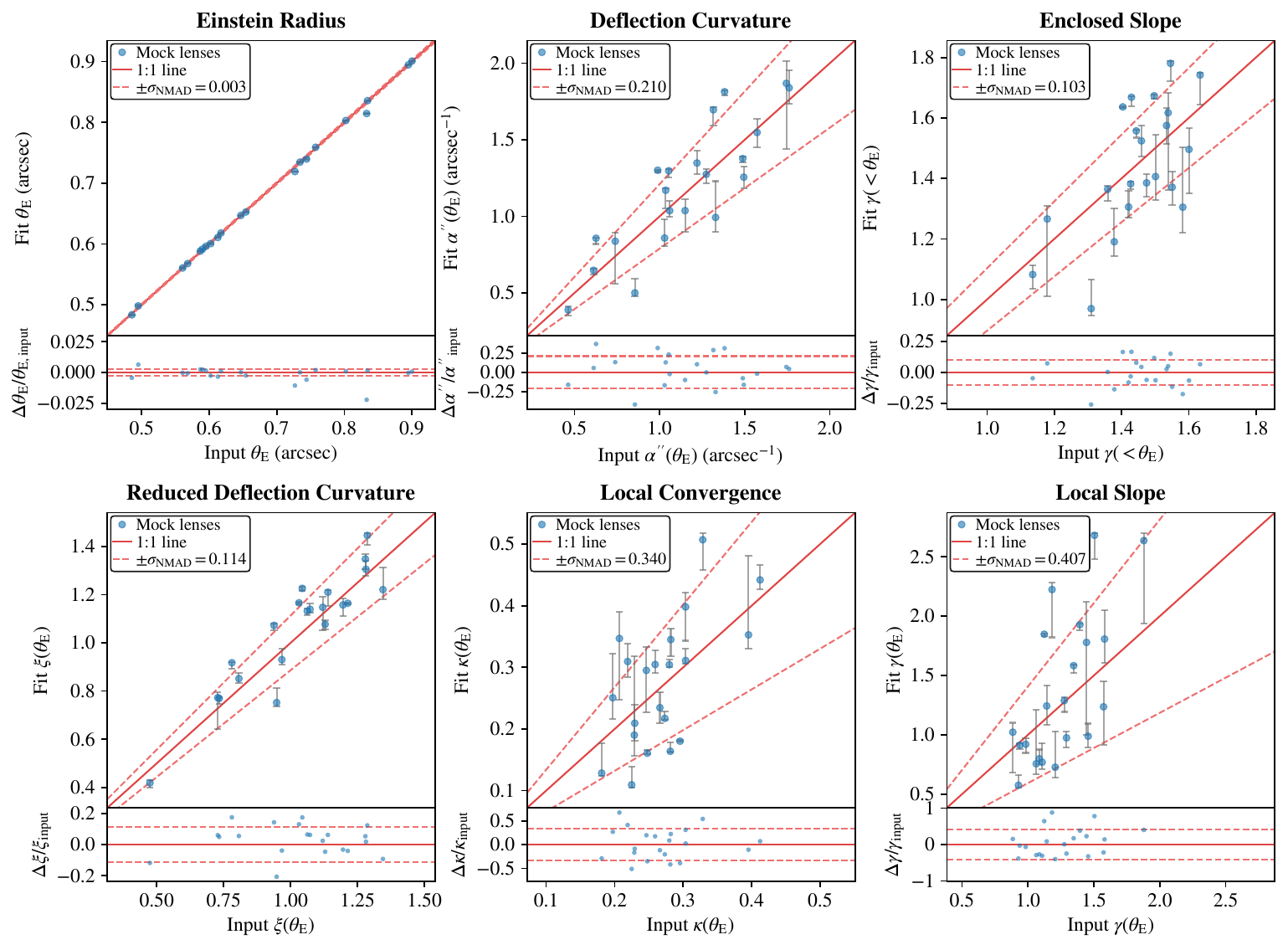}
    \caption{
    Recovered versus  input effective Einstein radii and circularised
    radial-structure diagnostics for the 21 systems with successful
    decomposed fits. From left to right, the top row shows the effective
    Einstein radius \(\theta_{\rm E}\), the second radial derivative of the
    circularised deflection profile \(\alpha''(\theta_{\rm E})\), and the
    global projected mass-weighted slope
    \(\gamma(<\theta_{\rm E})\). The bottom row shows the dimensionless
    mass-sheet-invariant reduced deflection curvature
    \(\xi(\theta_{\rm E})\), the local
    convergence \(\kappa(\theta_{\rm E})\), and the local projected
    mass-weighted slope \(\gamma(\theta_{\rm E})\), evaluated over
    \([0.95\theta_{\rm E},1.05\theta_{\rm E}]\). Fitted quantities are
    evaluated for every posterior draw at its corresponding effective
    Einstein radius; points show the posterior medians and vertical
    bars span the 16th to 84th percentiles. Lower subpanels show the
    signed fractional residuals used to calculate \(b_q\) and
    \(\sigma_{\rm NMAD}\), with
     $\Delta_q=(q_{\rm fit}-q_{\rm input})/q_{\rm input}$. Solid
    lines mark the one-to-one relation and zero residual, while dashed
    lines indicate the zero-centred \(\pm\sigma_{\rm NMAD}\) bounds.
    }
    \label{fig:radial_structure}
\end{figure*}

Fig.~\ref{fig:radial_structure} summarises the six recovered
quantities. The  input Einstein radii span approximately
0.5--0.9\arcsec. Across the 21 successful fits, the Einstein-radius residual
distribution has a median bias of \(\ensuremath{+0.01\%}\) and
\(\sigma_{\rm NMAD}=\ensuremath{0.3\%}\). The global slope and
reduced deflection curvature \(\xi\) have median biases of \(\ensuremath{+0.4\%}\) and
\(\ensuremath{+5.4\%}\), with \(\sigma_{\rm NMAD}\) values of \(\ensuremath{10.3\%}\) and \(\ensuremath{11.4\%}\),
respectively.

The remaining local diagnostics have larger \(\sigma_{\rm NMAD}\) values.
The deflection-curvature, local-convergence, and local-slope residuals have
median biases of \(\ensuremath{+4.5\%}\), \(\ensuremath{+2.4\%}\), and \(\ensuremath{+1.0\%}\), with
\(\sigma_{\rm NMAD}\) values of \(\ensuremath{21.0\%}\), \(\ensuremath{34.0\%}\), and \(\ensuremath{40.7\%}\),
respectively.       
    Because all six median biases are much smaller than their corresponding \(\sigma_{\rm NMAD}\) values, the recovered circularised profiles show no systematic steepening or shallowing near \(\theta_{\rm E}\), and differences in recovery precision are governed primarily by system-to-system scatter rather than by common directional offsets.

The six diagnostics separate into global or enclosed quantities and
local profile quantities. The former---\(\theta_{\rm E}\) and
\(\gamma(<\theta_{\rm E})\)---have the smallest NMAD scatters. Among the
local diagnostics, \(\xi\) has the smallest NMAD scatter and the
narrow-annulus slope \(\gamma(\theta_{\rm E})\) the largest. We discuss
the origin and interpretation of these differences in
Section~\ref{subsec:recovery_hierarchy}.

\subsection{Foreground-light residuals and true-light diagnostic}
\label{subsec:light_residuals}

The total aperture mass is recovered more accurately than the stellar and
dark-matter masses. In the final model, the fitted MGE light profile also
provides the spatial template for the stellar mass. Errors in the MGE fit can
therefore affect the inferred division between stellar and dark matter.

\begin{figure*}
    \centering
    \includegraphics[width=0.79\textwidth]{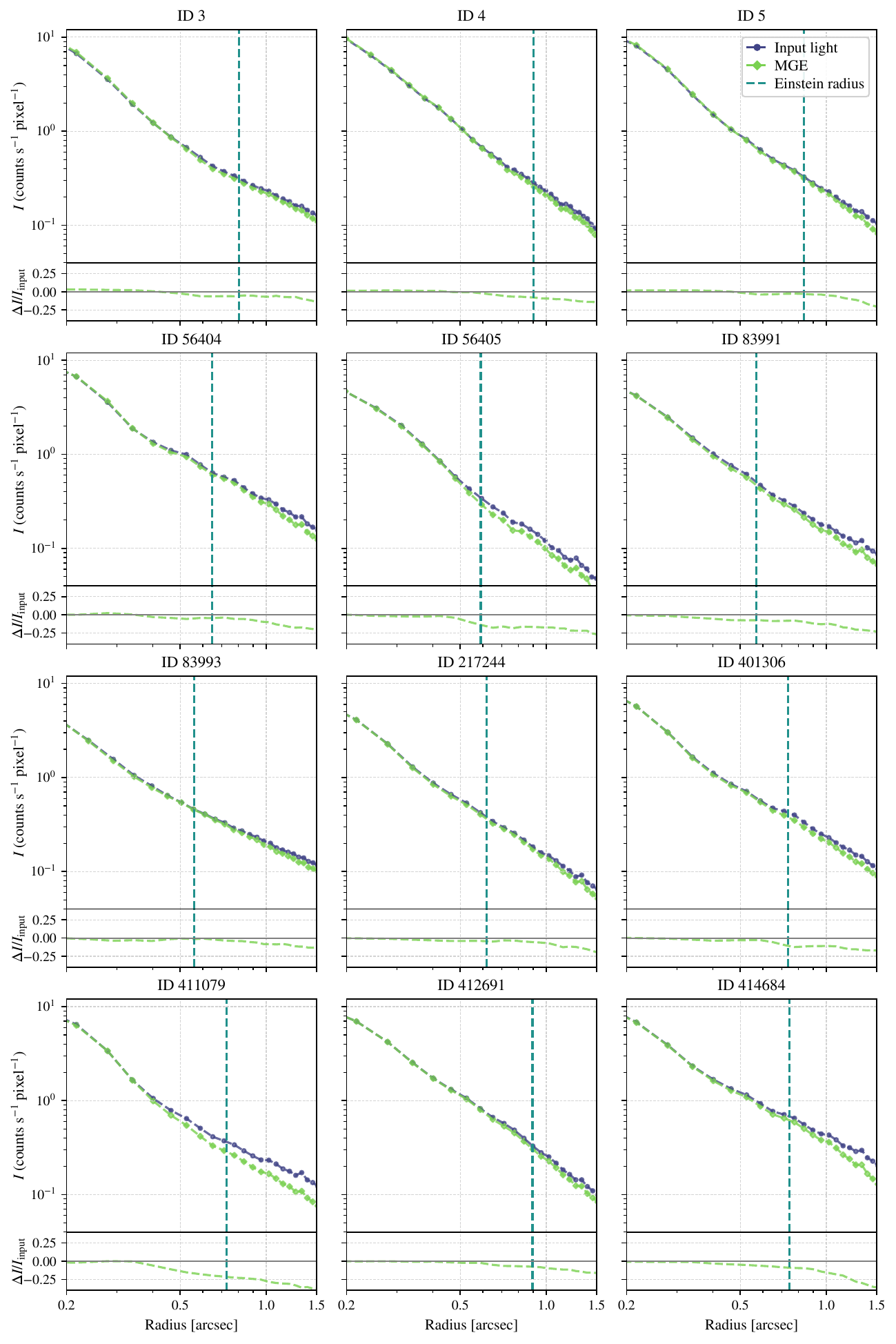}
    \caption{
    Radial surface-brightness profiles for 12 selected systems from the
    21-system sample:  input luminosity distribution (blue) and best-fit MGE
    models (green).
    Cyan dashed vertical lines mark the Einstein radius
    $\theta_{\rm E}$ of each system.  Lower panels show fractional
    residuals using the same  fit-minus-input convention as the other
    recovery figures:
     $\Delta I/I_{\rm input}=(I_{\rm fit}-I_{\rm input})/I_{\rm input}$.
    }
    \label{fig:mge}
\end{figure*}

Fig.~\ref{fig:mge} compares the fitted and  input radial light profiles for 12
systems from the 21-system successful-fit sample. The MGE reproduces the
overall profile in most cases, but differences of 10--40\% remain at some
radii. These differences are generally small near the centre and become
increasingly negative at larger radii, indicating that the MGE tends to
underestimate the outer light. Errors near \(\theta_{\rm E}\) are especially
relevant because the foreground light overlaps the lensed arcs in this region.

\begingroup
\setlength{\parskip}{0pt}
The MGE residuals motivate a complementary test in which the foreground light
is taken directly from the input. We therefore use the true-light diagnostic
described in Section~\ref{subsec:true_light_control}. In this diagnostic, the
exact input light distribution is subtracted from each mock image and used as
the spatial template for the stellar mass, while the global mass-to-light
ratio remains free. For computational efficiency, the diagnostic uses a
parametric source model rather than the pixelised source model adopted in the
main analysis. Across all 35 systems, the median biases in the total, stellar,
and dark-matter masses are \(\ensuremath{-0.11\%}\), \(\ensuremath{+2.1\%}\), and \(\ensuremath{-13.7\%}\),
respectively. The corresponding \(\sigma_{\rm NMAD}\) values are \(0.1\%\),
\(\ensuremath{0.8\%}\), and \(\ensuremath{6.3\%}\) (Fig.~\ref{fig:error_sources}). The smaller biases
and \(\sigma_{\rm NMAD}\) values for the stellar and dark-matter masses,
together with the MGE residuals in the main analysis, indicate that accurate
lens-light modelling is important for a reliable stellar--dark-matter
decomposition.

\begin{figure*}
    \centering
    \includegraphics[width=0.9\textwidth]{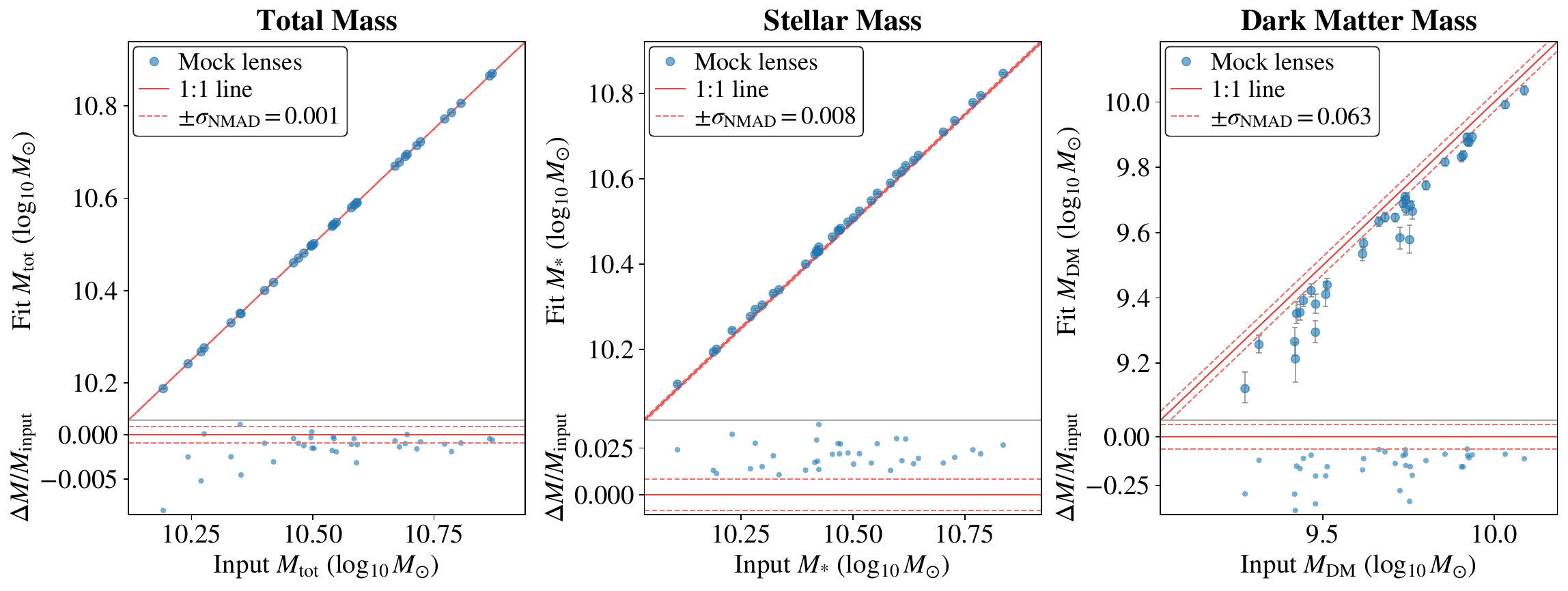}
    \caption{
    True-light diagnostic in which the  exact input light distribution is used 
    directly as the stellar-mass template. The panels compare the recovered and  input
    projected total, stellar, and dark-matter masses for all 35 systems.
    For each system, the masses
    are evaluated for every posterior draw; points show the posterior medians
    and vertical bars span the 16th to 84th percentiles. Bottom subpanels
    show the signed fractional residuals used to calculate \(b_q\) and
    \(\sigma_{\rm NMAD}\). Solid lines mark the one-to-one relation and
    zero residual; dashed lines indicate the zero-centred
    \(\pm\sigma_{\rm NMAD}\) bounds.
    }
    \label{fig:error_sources}
\end{figure*}


Fig.~\ref{fig:mge_limit} shows  two examples of the 14 systems not included in the
component-recovery statistics. Their coherent residuals along the disc plane
show where the adopted finite MGE basis fails to represent the radially
extended, vertically thin foreground light sufficiently well for the final
decomposed fit.

\begin{figure*}
    \centering
    \includegraphics[width=0.9\textwidth]{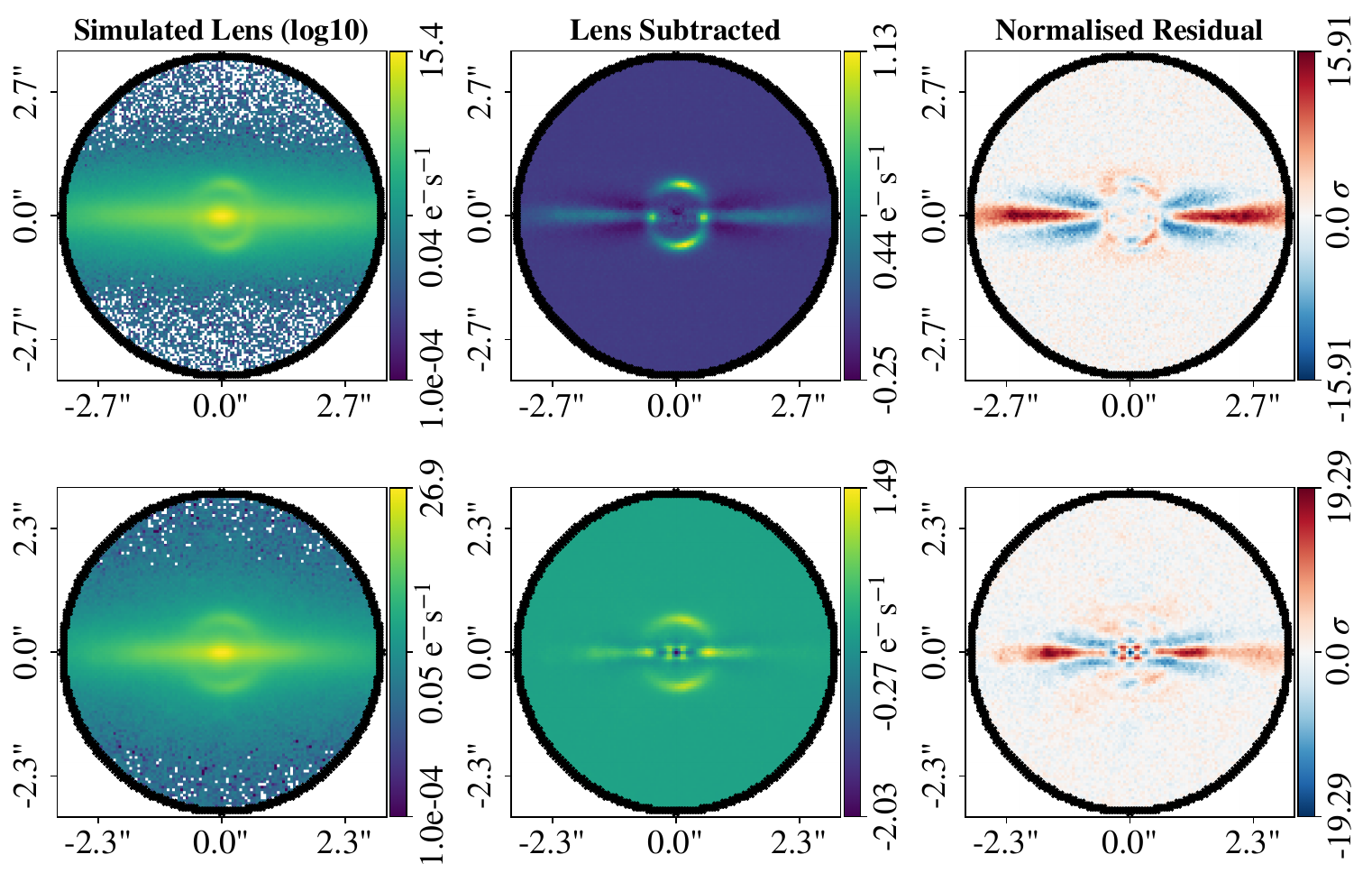}
    \caption{
    Two examples of edge-on disc systems excluded from the component-recovery
    analysis, shown in the upper and lower rows. From left to right, 
    each row shows the simulated lens image, the lens-light-subtracted image, and 
    the normalised residuals. Coherent residuals aligned with the disc plane highlight 
    the foreground-light subtraction mismatch that led to the exclusion of these systems 
    from the component-recovery analysis.
    }
    \label{fig:mge_limit}
\end{figure*}

\endgroup

\section{Discussion}
\label{sec:discussion}

\subsection{Global and local radial information}
\label{subsec:recovery_hierarchy}

Local radial quantities evaluated near the lensed arcs are not necessarily
recovered more accurately than the global projected mass-weighted slope
measured over the full Einstein aperture. The image configuration
directly fixes the Einstein radius
and the total projected mass inside it, so both quantities are recovered with
little scatter. In contrast, recovering the shape of the radial mass profile
is more model dependent. In particular, the global slope
\(\gamma(<\theta_{\rm E})\) is recovered substantially better than the
narrow-annulus local slope \(\gamma(\theta_{\rm E})\), with NMAD scatters of
\ensuremath{10.3\%} and \ensuremath{40.7\%}, respectively. The apparent locality of the lensing
constraints therefore does not guarantee that a locally defined profile
quantity will be recovered most accurately.

At first glance, it may seem puzzling that the global slope is recovered more accurately than the local slope, since strong-lensing constraints are theoretically localised near the Einstein ring. However, there is no real contradiction. As demonstrated by \citet{Cao2022EPLSystematics} when fitting elliptical power-law models to complex mock lenses, a mass-sheet-like transformation under mass-model mismatch can preserve both the lensing observables and the total enclosed mass inside \(\theta_{\rm E}\), yet still substantially twist the local radial profile near the critical curve. Consequently, the local slope evaluated in a narrow annulus around \(\theta_{\rm E}\) is much more sensitive to model mismatch than the global mass-weighted slope, which averages out local perturbations over the full Einstein aperture.

The reduced deflection curvature \(\xi\) adds a complementary perspective.
Its \ensuremath{11.4\%} NMAD scatter is much smaller than those of \(\alpha''\) and
\(\kappa\) individually, as expected because \(\xi\), rather than either
constituent term, is the mass-sheet-invariant combination constrained by the
local lens mapping \citep{Kochanek2019}. Nevertheless, \(\xi\) is recovered
only comparably to the global slope, whose NMAD scatter is slightly smaller.
This is not a contradiction: a quantity that is identifiable within the local
lens mapping need not be the quantity recovered most accurately relative to
the  input mass distribution when the fitted model is imperfect. Moreover, our
diagnostics use circularised profiles for intrinsically flattened edge-on
discs, so azimuthal averaging and the numerical sensitivity of local
derivatives may also contribute to their scatter. The robust interpretation
is therefore that \(\xi\) is the most accurately recovered local diagnostic
in these mocks, whereas the global slope is recovered with comparable
accuracy.

\subsection{Stellar--dark-matter decomposition and foreground-light modelling}
\label{subsec:light_discussion}

The accurate recovery of the total mass enclosed within the Einstein aperture does 
not translate into an equally robust stellar--dark-matter decomposition. As these apertures 
are predominantly stellar-dominated, the dark matter mass is inferred from subtracting the 
dominant stellar mass from the well-constrained total mass. Through this subtraction, 
modest systematic or statistical errors in the stellar component propagate into substantial 
fractional errors in the subdominant dark matter. This error propagation naturally explains 
the oppositely directed biases between the components and the much larger scatter in the 
inferred dark matter mass, without implying any degradation in the total aperture mass constraint.

Foreground light modelling enters this mass decomposition through two distinct pathways. 
First, the MGE separates the foreground emission from the lensed source; second, when 
scaled by the global mass-to-light ratio, it dictates the spatial distribution of the 
stellar mass. Consequently, any mismatch in the MGE fit propagates into both the reconstructed 
source and the stellar mass template. Similar systematic dependencies on lens-light subtraction 
have been reported in automated analyses of HST lenses \citep{Nightingale2024SubhaloScan}.

The structured MGE residuals in the main analysis, together with the
improved component recovery in the true-light diagnostic, indicate that
accurate foreground-light modelling is important for reliable
stellar--dark-matter decomposition. The 14 unsuccessful systems further mark a practical limit of
the adopted workflow: their disc-aligned residuals show that the finite MGE
basis cannot adequately represent some radially extended, vertically thin
light distributions. This limitation could be addressed through either more
physically motivated inclined-disc profiles, which model the radial and
vertical structure separately \citep[e.g.][]{Erwin2015Imfit}, or non-parametric
foreground-light components based on information field theory
\citep[e.g.][]{Rustig2024LensCharm,Galan2024ElAnzuelo,Edenhofer2024NIFTyre}.

\subsection{Limitations and observational prospects}
\label{subsec:limitations_implications}

By design, our mock suite serves as a controlled, simplified benchmark. 
The synthetic data and inference framework share matched physical assumptions, 
namely a spatially invariant stellar mass-to-light ratio and an identical spherical NFW mass--concentration relation. 
Moreover, all mock galaxies are viewed strictly edge-on, with our summary statistics 
conditioned exclusively on the 21 successfully decomposed systems. The
resulting precision applies to this controlled sample and is not intended to
represent the broader observed population of disc lenses, which spans wider
ranges of morphology and inclination
\citep{Treu2011SWELLS,Brewer2012SWELLS,AcevedoBarroso2025}.

Real lenses introduce additional couplings. Dust and stellar-population
gradients can make the observed light an imperfect tracer of stellar mass, so
a single global \(M/L\) may bias the inferred decomposition even when the
surface brightness is fitted accurately \citep{Liang2025MLGradient}. Halo
shape, substructure, and more complex source morphology add further departures
from the matched assumptions used here.

Applying the workflow to the SWELLS data would extend the analysis from
controlled mocks to real disc lenses, but it would not provide a direct test of
recovery accuracy because their true mass distributions are unknown. Instead,
SWELLS can reveal coherent residuals and sensitivity of the inferred quantities
to alternative light and mass models across real morphologies, dust structures,
and inclinations. Multi-band imaging can help separate foreground lens light,
dust attenuation, and lensed-source emission, while available stellar or gas
kinematics can provide an external consistency check on the imaging-based
decomposition \citep{Treu2011SWELLS,Dutton2011SWELLS}. Larger UNIONS and Euclid
samples could characterize how often structured residuals or strong model
dependence occur, and whether they are more common in highly inclined,
disc-dominated, or dust-rich lenses
\citep{AcevedoBarroso2025,EuclidWalmsley2026}. These observational diagnostics
can expose failure modes, but they cannot establish recovery accuracy, which
must be calibrated using controlled mocks with known truth.

\section{Conclusion}
\label{sec:conclusion}

Using controlled TNG50 mock lenses, we find that imaging of edge-on disc lenses
robustly recovers the Einstein radius, enclosed total mass, and global projected
mass-weighted slope. The global slope is recovered substantially more accurately
than the local slope around the Einstein radius and slightly more accurately
than the reduced deflection curvature \(\xi\). Although \(\xi\) remains the most
robust local diagnostic, these results show that proximity to the lensed arcs
does not guarantee greater accuracy relative to the  input mass distribution.

The stellar--dark-matter decomposition is more sensitive to foreground-light
modelling. Component recovery improves in the true-light diagnostic, while the
adopted finite MGE basis fails for 14 of the 35 systems. Our quantitative
conclusions therefore apply to the 21 successfully decomposed, exactly edge-on
systems under largely matched mass-model assumptions. Extending these
inferences to real lenses requires explicit checks for structured residuals and
sensitivity to the adopted light, source, and mass models.

In future work, we will apply this modelling framework to real observations of edge-on disc lenses, including large samples expected from wide-field space-based surveys such as \textit{Euclid} and CSST.

\section*{Acknowledgements}
We acknowledge support from the National Natural Science Foundation of
China (Nos. 12588202 and 12573007), the China Manned Space Program
(Nos. CMS-CSST-2025-A03 and CMS-CSST-2025-A20), and the CAS Project for
Young Scientists in Basic Research (No. YSBR-062).
XYC acknowledges the support of the National Natural Science Foundation of China (No.\ 12303006).
We thank the IllustrisTNG team for making the simulation data publicly
available.
The IllustrisTNG simulations were undertaken with compute time awarded by
the Gauss Centre for Supercomputing (GCS) under GCS Large-Scale Projects
GCS-ILLU and GCS-DWAR on the GCS share of the supercomputer Hazel Hen at
the High Performance Computing Center Stuttgart (HLRS), as well as on the
machines of the Max Planck Computing and Data Facility (MPCDF) in Garching,
Germany.

\section*{Data Availability}
\begingroup
\urlstyle{same}
\setlength{\fboxsep}{0pt}
The TNG50 simulation data are publicly available at
\url{https://www.tng-project.org/data/} \citep{Nelson2019TNG}.
The code used in this work is hosted at
\href{https://github.com/fangzheng09/disk-lens-model}{\strut https://}%
\allowbreak\href{https://github.com/fangzheng09/disk-lens-model}{\strut github.com/}%
\allowbreak\href{https://github.com/fangzheng09/disk-lens-model}{\strut fangzheng09/}%
\allowbreak\href{https://github.com/fangzheng09/disk-lens-model}{\strut disk-lens-model}.
The mock lens images and modelling outputs are available upon
reasonable request to the corresponding author.
\par
\endgroup

\bibliographystyle{mnras}
\bibliography{references}

@article{2010PhR...495...33B,
  author = {Benson, Andrew J.},
  title = {Galaxy Formation Theory},
  journal = {Physics Reports},
  year = {2010},
  volume = {495},
  number = {2--3},
  pages = {33--86},
  doi = {10.1016/j.physrep.2010.06.001},
  eprint = {1006.5394},
  archivePrefix = {arXiv}
}

@article{AcevedoBarroso2025,
  author = {Acevedo Barroso, J. A. and Cl{\'e}ment, B. and Courbin, F. and Gavazzi, R. and Lemon, C. and Rojas, K. and Scott, D. and Gwyn, S. and Hammer, F. and Hudson, M. J. and Magnier, E. A.},
  title = {Searching for strong lensing by late-type galaxies in {UNIONS}},
  journal = {Astronomy \& Astrophysics},
  volume = {706},
  pages = {A146},
  year = {2026},
  doi = {10.1051/0004-6361/202554555},
  archivePrefix = {arXiv},
  eprint = {2503.10610},
  primaryClass = {astro-ph.GA}
}

@article{Allgood2006Shapes,
  author = {Allgood, Brandon and Flores, Ricardo A. and Primack, Joel R. and Kravtsov, Andrey V. and Wechsler, Risa H. and Faltenbacher, Andreas and Bullock, James S.},
  title = {The shape of dark matter haloes: dependence on mass, redshift, radius and formation},
  journal = {Monthly Notices of the Royal Astronomical Society},
  volume = {367},
  number = {4},
  pages = {1781--1796},
  year = {2006},
  doi = {10.1111/j.1365-2966.2006.10094.x}
}

@article{Barnabe2012SWELLS,
  author = {Barnab{\`e}, Matteo and Dutton, Aaron A. and Marshall, Philip J. and Auger, Matthew W. and Brewer, Brendon J. and Treu, Tommaso and Bolton, Adam S. and Koo, David C. and Koopmans, L{\'e}on V. E.},
  title = {The {SWELLS} survey -- {IV}. Precision measurements of the stellar and dark matter distributions in a spiral lens galaxy},
  journal = {Monthly Notices of the Royal Astronomical Society},
  year = {2012},
  volume = {423},
  number = {2},
  pages = {1073--1088},
  doi = {10.1111/j.1365-2966.2012.20934.x},
  eprint = {1201.1692},
  archivePrefix = {arXiv}
}

@article{Brewer2012SWELLS,
  author = {Brewer, Brendon J. and Dutton, Aaron A. and Treu, Tommaso and Auger, Matthew W. and Marshall, Philip J. and Barnab{\`e}, Matteo and Bolton, Adam S. and Koo, David C. and Koopmans, L{\'e}on V. E.},
  title = {The {SWELLS} survey -- III. Disfavouring `heavy' initial mass functions for spiral lens galaxies},
  journal = {Monthly Notices of the Royal Astronomical Society},
  volume = {422},
  number = {4},
  pages = {3574--3590},
  year = {2012},
  doi = {10.1111/j.1365-2966.2012.20870.x}
}

@article{Cao2022EPLSystematics,
  author = {Cao, Xiaoyue and Li, Ran and Nightingale, James W. and others},
  title = {Systematic errors induced by the elliptical power-law model in galaxy--galaxy strong lens modeling},
  journal = {Research in Astronomy and Astrophysics},
  volume = {22},
  pages = {025014},
  year = {2022},
  doi = {10.1088/1674-4527/ac3f2b}
}

@article{Cao2023CSSTForecast,
  author = {Cao, Xiaoyue and Li, Ran and Li, Nan and Li, Rui and Chen, Yun and Ding, Keyi and Shan, Huanyuan and Zhan, Hu and Zhang, Xin and Du, Wei and Cao, Shuo},
  title = {{CSST} Strong Lensing Preparation: Forecasting the galaxy--galaxy strong lensing population for the {China Space Station Telescope}},
  journal = {Monthly Notices of the Royal Astronomical Society},
  volume = {533},
  number = {2},
  pages = {1960--1975},
  year = {2024},
  doi = {10.1093/mnras/stae1865},
  archivePrefix = {arXiv},
  eprint = {2312.06239},
  primaryClass = {astro-ph.CO}
}

@article{Cappellari2002MGE,
  author = {Cappellari, Michele},
  title = {Efficient multi-Gaussian expansion of galaxies},
  journal = {Monthly Notices of the Royal Astronomical Society},
  volume = {333},
  number = {2},
  pages = {400--410},
  year = {2002},
  doi = {10.1046/j.1365-8711.2002.05412.x},
  archivePrefix = {arXiv},
  eprint = {astro-ph/0201430}
}

@article{Collett2015,
  author = {Collett, Thomas E.},
  title = {The population of galaxy--galaxy strong lenses in forthcoming optical imaging surveys},
  journal = {The Astrophysical Journal},
  volume = {811},
  pages = {20},
  year = {2015},
  doi = {10.1088/0004-637X/811/1/20}
}

@article{Courteau2014GalaxyMasses,
  author = {Courteau, St{\'e}phane and Cappellari, Michele and de Jong, Roelof S. and Dutton, Aaron A. and Emsellem, Eric and Hoekstra, Henk and Koopmans, L. V. E. and Mamon, Gary A. and Maraston, Claudia and Treu, Tommaso and Widrow, Lawrence M.},
  title = {Galaxy Masses},
  journal = {Reviews of Modern Physics},
  year = {2014},
  volume = {86},
  number = {1},
  pages = {47--119},
  doi = {10.1103/RevModPhys.86.47},
  eprint = {1309.3276},
  archivePrefix = {arXiv}
}

@article{Davis1985,
  author = {Davis, Marc and Efstathiou, George and Frenk, Carlos S. and White, Simon D. M.},
  title = {The evolution of large-scale structure in a universe dominated by cold dark matter},
  journal = {The Astrophysical Journal},
  volume = {292},
  pages = {371--394},
  year = {1985},
  doi = {10.1086/163168}
}

@article{Ding2021TDLMC,
  author = {Ding, Xuheng and Treu, Tommaso and Birrer, Simon and others},
  title = {Time delay lens modelling challenge},
  journal = {Monthly Notices of the Royal Astronomical Society},
  volume = {503},
  number = {1},
  pages = {1096--1123},
  year = {2021},
  doi = {10.1093/mnras/stab484},
  archivePrefix = {arXiv},
  eprint = {2006.08619},
  primaryClass = {astro-ph.CO}
}

@article{Dutton2011SWELLS,
  author = {Dutton, Aaron A. and Brewer, Brendon J. and Marshall, Philip J. and Auger, Matthew W. and Treu, Tommaso and Koo, David C. and Bolton, Adam S. and Holden, Bradford P. and Koopmans, L{\'e}on V. E.},
  title = {The {SWELLS} survey -- II. Breaking the disc--halo degeneracy in the spiral galaxy gravitational lens {SDSS J2141-0001}},
  journal = {Monthly Notices of the Royal Astronomical Society},
  volume = {417},
  pages = {1621--1642},
  year = {2011},
  doi = {10.1111/j.1365-2966.2011.18706.x}
}

@article{DuttonTreu2014,
  author = {Dutton, Aaron A. and Treu, Tommaso},
  title = {The bulge--halo conspiracy in massive elliptical galaxies: implications for the stellar initial mass function and halo response to baryonic processes},
  journal = {Monthly Notices of the Royal Astronomical Society},
  volume = {438},
  number = {4},
  pages = {3594--3602},
  year = {2014},
  doi = {10.1093/mnras/stt2489}
}

@article{Edenhofer2024NIFTyre,
  author = {Edenhofer, Gordian and Frank, Philipp and Roth, Jakob and Leike, Reimar H. and Guerdi, Massin and Scheel-Platz, Lukas I. and Guardiani, Matteo and Eberle, Vincent and Westerkamp, Margret and En{\ss}lin, Torsten A.},
  title = {Re-Envisioning Numerical Information Field Theory ({NIFTy.re}): A Library for Gaussian Processes and Variational Inference},
  journal = {Journal of Open Source Software},
  year = {2024},
  volume = {9},
  number = {98},
  pages = {6593},
  doi = {10.21105/joss.06593},
  url = {https://joss.theoj.org/papers/10.21105/joss.06593}
}

@article{Erwin2015Imfit,
  author = {Erwin, Peter},
  title = {{Imfit}: A Fast, Flexible New Program for Astronomical Image Fitting},
  journal = {The Astrophysical Journal},
  year = {2015},
  volume = {799},
  number = {2},
  pages = {226},
  doi = {10.1088/0004-637X/799/2/226},
  eprint = {1408.1097},
  archivePrefix = {arXiv}
}

@article{Etherington2022NoLensLeftBehind,
  author = {Etherington, Amy and Nightingale, James W. and Massey, Richard and Cao, XiaoYue and Robertson, Andrew and Amorisco, Nicola C. and Amvrosiadis, Aristeidis and Cole, Shaun and Frenk, Carlos S. and He, Qiuhan and Li, Ran and Tam, Sut-Ieng},
  title = {Automated galaxy--galaxy strong lens modelling: no lens left behind},
  journal = {Monthly Notices of the Royal Astronomical Society},
  volume = {517},
  number = {3},
  pages = {3275--3302},
  year = {2022},
  doi = {10.1093/mnras/stac2639},
  archivePrefix = {arXiv},
  eprint = {2202.09201}
}

@article{EuclidWalmsley2026,
  author = {{Euclid Collaboration} and Walmsley, M. and Holloway, P. and Lines, N. E. P. and Rojas, K. and Collett, T. E. and Verma, A. and Li, T. and others},
  title = {{Euclid} Quick Data Release ({Q1}): {XXVI}. The Strong Lensing Discovery Engine {A} -- System overview and first lens sample},
  journal = {Astronomy \& Astrophysics},
  year = {2026},
  volume = {711},
  pages = {A26},
  doi = {10.1051/0004-6361/202555141},
  eprint = {2503.15324},
  archivePrefix = {arXiv}
}

@article{Galan2024ElAnzuelo,
  author = {Galan, A. and Caminha, G. B. and Knollm{\"u}ller, J. and Roth, J. and Suyu, S. H.},
  title = {El Gordo needs El Anzuelo: Probing the structure of cluster members with multi-band extended arcs in {JWST} data},
  journal = {Astronomy \& Astrophysics},
  year = {2024},
  volume = {689},
  pages = {A304},
  doi = {10.1051/0004-6361/202449876},
  eprint = {2402.18636},
  archivePrefix = {arXiv}
}

@article{He2024MGE,
  author = {He, Qiuhan and Nightingale, James W. and Amvrosiadis, Aristeidis and Robertson, Andrew and Cole, Shaun and Frenk, Carlos S. and Massey, Richard and Li, Ran and Cao, Xiaoyue and Lange, Samuel C. and Fran{\c{c}}a, Jo{\~a}o Paulo C.},
  title = {Unveiling lens light complexity with a novel Multi-Gaussian Expansion approach for strong gravitational lensing},
  journal = {Monthly Notices of the Royal Astronomical Society},
  volume = {532},
  number = {2},
  pages = {2441--2462},
  year = {2024},
  doi = {10.1093/mnras/stae1577},
  archivePrefix = {arXiv},
  eprint = {2403.16253},
  primaryClass = {astro-ph.GA}
}

@article{Kochanek2019,
  author = {Kochanek, C. S.},
  title = {Overconstrained gravitational lens models and the {Hubble} constant},
  journal = {Monthly Notices of the Royal Astronomical Society},
  volume = {493},
  pages = {1725--1735},
  year = {2020},
  doi = {10.1093/mnras/staa344},
  archivePrefix = {arXiv},
  eprint = {1911.05083}
}

@article{Kormann1994,
  author = {Kormann, R. and Schneider, P. and Bartelmann, M.},
  title = {Isothermal elliptical gravitational lens models},
  journal = {Astronomy and Astrophysics},
  volume = {284},
  pages = {285--299},
  year = {1994}
}

@article{Lange2023Nautilus,
  author = {Lange, Johannes U.},
  title = {{nautilus}: boosting Bayesian importance nested sampling with deep learning},
  journal = {Monthly Notices of the Royal Astronomical Society},
  volume = {525},
  number = {2},
  pages = {3181--3194},
  year = {2023},
  doi = {10.1093/mnras/stad2441},
  archivePrefix = {arXiv},
  eprint = {2306.16923}
}

@article{Liang2025MLGradient,
  author = {Liang, Yan and Xu, Dandan and Sluse, Dominique and Sonnenfeld, Alessandro and Shu, Yiping},
  title = {Modelling biases from constant stellar mass-to-light ratio assumption in galaxy dynamics and strong lensing},
  journal = {Monthly Notices of the Royal Astronomical Society},
  volume = {536},
  pages = {2672--2689},
  year = {2025},
  doi = {10.1093/mnras/stae2744},
  archivePrefix = {arXiv},
  eprint = {2311.07442}
}

@article{Lines2025Euclid,
  author = {Lines, Natalie E. P. and Li, Tian and Collett, Thomas E. and Holloway, Philip and Nightingale, James W. and Rojas, Karina and Verma, Aprajita and Walmsley, Mike},
  title = {The revolution in strong lensing discoveries from {Euclid}},
  journal = {Nature Astronomy},
  year = {2025},
  volume = {9},
  pages = {1116--1122},
  doi = {10.1038/s41550-025-02616-5},
  eprint = {2508.14624},
  archivePrefix = {arXiv}
}

@article{Ludlow2016,
  author = {Ludlow, Aaron D. and Bose, Sownak and Angulo, Raul E. and Wang, Liang and Hellwing, Wojciech A. and Navarro, Julio F. and Cole, Shaun and Frenk, Carlos S.},
  title = {The mass--concentration--redshift relation of cold and warm dark matter haloes},
  journal = {Monthly Notices of the Royal Astronomical Society},
  volume = {460},
  number = {2},
  pages = {1214--1232},
  year = {2016},
  doi = {10.1093/mnras/stw1046}
}

@article{MoMaoWhite1998,
  author = {Mo, H. J. and Mao, Shude and White, Simon D. M.},
  title = {The formation of galactic discs},
  journal = {Monthly Notices of the Royal Astronomical Society},
  year = {1998},
  volume = {295},
  number = {2},
  pages = {319--336},
  doi = {10.1046/j.1365-8711.1998.01227.x}
}

@article{Monaghan1992SPH,
  author = {Monaghan, J. J.},
  title = {Smoothed Particle Hydrodynamics},
  journal = {Annual Review of Astronomy and Astrophysics},
  year = {1992},
  volume = {30},
  pages = {543--574},
  doi = {10.1146/annurev.aa.30.090192.002551}
}

@article{Mukherjee2018SEAGLE,
  author = {Mukherjee, Sampath and Koopmans, L{\'e}on V. E. and Metcalf, R. Benton and Tessore, Nicolas and Tortora, Crescenzo and Schaller, Matthieu and Schaye, Joop and Crain, Robert A. and Vernardos, Georgios and Bellagamba, Fabio and Theuns, Tom},
  title = {{SEAGLE} -- I: A pipeline for simulating and modelling strong lenses from cosmological hydrodynamical simulations},
  journal = {Monthly Notices of the Royal Astronomical Society},
  volume = {479},
  pages = {4108--4125},
  year = {2018},
  doi = {10.1093/mnras/sty1741},
  archivePrefix = {arXiv},
  eprint = {1802.06629}
}

@article{Navarro1996,
  author = {Navarro, Julio F. and Frenk, Carlos S. and White, Simon D. M.},
  title = {The structure of cold dark matter halos},
  journal = {The Astrophysical Journal},
  volume = {462},
  pages = {563--575},
  year = {1996},
  doi = {10.1086/177173}
}

@article{Nelson2019TNG,
  author = {Nelson, Dylan and Springel, Volker and Pillepich, Annalisa and Rodriguez-Gomez, Vicente and Torrey, Paul and Genel, Shy and Vogelsberger, Mark and Pakmor, R{\"u}diger and Marinacci, Federico and Weinberger, Rainer and others},
  title = {The {IllustrisTNG} simulations: public data release},
  journal = {Computational Astrophysics and Cosmology},
  volume = {6},
  pages = {2},
  year = {2019},
  doi = {10.1186/s40668-019-0028-x}
}

@article{Nelson2019TNG50,
  author = {Nelson, Dylan and Pillepich, Annalisa and Springel, Volker and Pakmor, R{\"u}diger and Weinberger, Rainer and Genel, Shy and Torrey, Paul and Vogelsberger, Mark and Marinacci, Federico and Hernquist, Lars},
  title = {First results from the {TNG50} simulation: galactic outflows driven by supernovae and black hole feedback},
  journal = {Monthly Notices of the Royal Astronomical Society},
  year = {2019},
  volume = {490},
  number = {3},
  pages = {3234--3261},
  doi = {10.1093/mnras/stz2306},
  eprint = {1902.05554},
  archivePrefix = {arXiv}
}

@article{Nightingale2018AutoLens,
  author = {Nightingale, James W. and Dye, Simon and Massey, Richard J.},
  title = {{AutoLens}: automated modelling of a strong lens's light, mass and source},
  journal = {Monthly Notices of the Royal Astronomical Society},
  volume = {478},
  pages = {4738--4784},
  year = {2018},
  doi = {10.1093/mnras/sty1264}
}

@article{Nightingale2019GalaxyStructure,
  author = {Nightingale, James W. and Massey, Richard J. and Harvey, David R. and Cooper, Andrew P. and Etherington, Amy and Tam, Sut-Ieng and Hayes, Richard G.},
  title = {Galaxy structure with strong gravitational lensing: decomposing the internal mass distribution of massive elliptical galaxies},
  journal = {Monthly Notices of the Royal Astronomical Society},
  year = {2019},
  volume = {489},
  number = {2},
  pages = {2049--2068},
  doi = {10.1093/mnras/stz2220},
  eprint = {1901.07801},
  archivePrefix = {arXiv},
  primaryClass = {astro-ph.GA},
  url = {https://academic.oup.com/mnras/article/489/2/2049/5550746}
}

@article{Nightingale2021PyAutoFit,
  author = {Nightingale, James W. and Hayes, Richard G. and Griffiths, Matthew},
  title = {{PyAutoFit}: A Classy Probabilistic Programming Language for Model Composition and Fitting},
  journal = {Journal of Open Source Software},
  year = {2021},
  volume = {6},
  number = {58},
  pages = {2550},
  doi = {10.21105/joss.02550},
  url = {https://joss.theoj.org/papers/10.21105/joss.02550}
}

@article{Nightingale2021PyAutoLens,
  author = {Nightingale, James W. and Hayes, Richard G. and Kelly, Ashley and Amvrosiadis, Aristeidis and Etherington, Amy and He, Qiuhan and Li, Nan and Cao, XiaoYue and Frawley, Jonathan and Cole, Shaun and Enia, Andrea and Frenk, Carlos S. and Harvey, David R. and Li, Ran and Massey, Richard J. and Negrello, Mattia and Robertson, Andrew},
  title = {{PyAutoLens}: Open-source strong gravitational lensing},
  journal = {Journal of Open Source Software},
  volume = {6},
  number = {58},
  pages = {2825},
  year = {2021},
  doi = {10.21105/joss.02825},
  archivePrefix = {arXiv},
  eprint = {2106.01384}
}

@article{Nightingale2023PyAutoGalaxy,
  author = {Nightingale, James W. and Amvrosiadis, Aristeidis and Hayes, Richard G. and He, Qiuhan and Etherington, Amy and Cao, XiaoYue and Cole, Shaun and Frawley, Jonathan and Frenk, Carlos S. and Lange, Sam and Li, Ran and Massey, Richard J. and Negrello, Mattia and Robertson, Andrew},
  title = {{PyAutoGalaxy}: Open-Source Multiwavelength Galaxy Structure \& Morphology},
  journal = {Journal of Open Source Software},
  year = {2023},
  volume = {8},
  number = {81},
  pages = {4475},
  doi = {10.21105/joss.04475},
  url = {https://joss.theoj.org/papers/10.21105/joss.04475}
}

@article{Nightingale2024SubhaloScan,
  author = {Nightingale, James W. and He, Qiuhan and Cao, Xiaoyue and Amvrosiadis, Aristeidis and Etherington, Amy and Frenk, Carlos S. and Hayes, Richard G. and Robertson, Andrew and Cole, Shaun and Lange, Samuel and Li, Ran and Massey, Richard},
  title = {Scanning for dark matter subhaloes in {Hubble Space Telescope} imaging of 54 strong lenses},
  journal = {Monthly Notices of the Royal Astronomical Society},
  year = {2024},
  volume = {527},
  number = {4},
  pages = {10480--10506},
  doi = {10.1093/mnras/stad3694},
  eprint = {2209.10566},
  archivePrefix = {arXiv},
  primaryClass = {astro-ph.CO},
  url = {https://academic.oup.com/mnras/article/527/4/10480/7458645}
}

@article{NightingaleDye2015,
  author = {Nightingale, James W. and Dye, Simon},
  title = {Adaptive semi-linear inversion of strong gravitational lens imaging},
  journal = {Monthly Notices of the Royal Astronomical Society},
  volume = {452},
  number = {3},
  pages = {2940--2959},
  year = {2015},
  doi = {10.1093/mnras/stv1455}
}

@article{Pillepich2018TNGModel,
  author = {Pillepich, Annalisa and Springel, Volker and Nelson, Dylan and Genel, Shy and Naiman, Jill and Pakmor, R{\"u}diger and Hernquist, Lars and Torrey, Paul and Vogelsberger, Mark and Weinberger, Rainer and Marinacci, Federico},
  title = {Simulating galaxy formation with the {IllustrisTNG} model},
  journal = {Monthly Notices of the Royal Astronomical Society},
  volume = {473},
  pages = {4077--4106},
  year = {2018},
  doi = {10.1093/mnras/stx2656},
  archivePrefix = {arXiv},
  eprint = {1703.02970}
}

@article{Pillepich2019TNG50,
  author = {Pillepich, Annalisa and Nelson, Dylan and Springel, Volker and Pakmor, R{\"u}diger and Torrey, Paul and Weinberger, Rainer and Vogelsberger, Mark and Marinacci, Federico and Genel, Shy and van der Wel, Arjen and Hernquist, Lars},
  title = {First results from the {TNG50} simulation: the evolution of stellar and gaseous discs across cosmic time},
  journal = {Monthly Notices of the Royal Astronomical Society},
  volume = {490},
  pages = {3196--3233},
  year = {2019},
  doi = {10.1093/mnras/stz2338},
  archivePrefix = {arXiv},
  eprint = {1902.05553}
}

@article{Planck2016,
  author = {{Planck Collaboration} and Ade, P. A. R. and others},
  title = {Planck 2015 results. {XIII}. Cosmological parameters},
  journal = {Astronomy \& Astrophysics},
  volume = {594},
  pages = {A13},
  year = {2016},
  doi = {10.1051/0004-6361/201525830}
}

@article{Planck2020,
  author = {{Planck Collaboration}},
  title = {{Planck} 2018 results. {VI}. Cosmological parameters},
  journal = {Astronomy \& Astrophysics},
  year = {2020},
  volume = {641},
  pages = {A6},
  doi = {10.1051/0004-6361/201833910},
  eprint = {1807.06209},
  archivePrefix = {arXiv},
  primaryClass = {astro-ph.CO},
  url = {https://www.aanda.org/articles/aa/abs/2020/09/aa33910-18/aa33910-18.html}
}

@article{Price2007SPLASH,
  author = {Price, Daniel J.},
  title = {{SPLASH}: An Interactive Visualisation Tool for Smoothed Particle Hydrodynamics Simulations},
  journal = {Publications of the Astronomical Society of Australia},
  year = {2007},
  volume = {24},
  number = {3},
  pages = {159--173},
  doi = {10.1071/AS07022},
  eprint = {0709.0832},
  archivePrefix = {arXiv}
}

@article{Puchwein2009EinsteinRadius,
  author = {Puchwein, Ewald and Hilbert, Stefan},
  title = {Cluster strong lensing in the {Millennium} simulation: the effect of galaxies and structures along the line-of-sight},
  journal = {Monthly Notices of the Royal Astronomical Society},
  volume = {398},
  number = {3},
  pages = {1298--1308},
  year = {2009},
  doi = {10.1111/j.1365-2966.2009.15227.x},
  archivePrefix = {arXiv},
  eprint = {0904.0253},
  primaryClass = {astro-ph.GA}
}

@article{RodriguezGomez2015,
  author = {Rodriguez-Gomez, Vicente and Genel, Shy and Vogelsberger, Mark and Sijacki, Debora and Pillepich, Annalisa and Sales, Laura V. and Torrey, Paul and Snyder, Gregory and Nelson, Dylan and Springel, Volker and Ma, Chung-Pei and Hernquist, Lars},
  title = {The merger rate of galaxies in the {Illustris} simulation: a comparison with observations and semi-empirical models},
  journal = {Monthly Notices of the Royal Astronomical Society},
  volume = {449},
  pages = {49--64},
  year = {2015},
  doi = {10.1093/mnras/stv264}
}

@article{Rustig2024LensCharm,
  author = {R{\"u}stig, Julian and Guardiani, Matteo and Roth, Jakob and Frank, Philipp and En{\ss}lin, Torsten},
  title = {Introducing {LensCharm}: A charming Bayesian strong lensing reconstruction framework},
  journal = {Astronomy \& Astrophysics},
  year = {2024},
  volume = {682},
  pages = {A146},
  doi = {10.1051/0004-6361/202348256},
  url = {https://www.aanda.org/articles/aa/full_html/2024/02/aa48256-23/aa48256-23.html}
}

@misc{STScIACSZeropoints,
  author = {{Space Telescope Science Institute}},
  title = {{ACS} Zeropoints},
  year = {2026},
  howpublished = {Hubble Space Telescope instrument calibration documentation},
  url = {https://www.stsci.edu/hst/instrumentation/acs/data-analysis/zeropoints},
  note = {Last updated 2026 April 16; accessed 2026 October 23}
}

@article{Sales2010Feedback,
  author = {Sales, Laura V. and Navarro, Julio F. and Schaye, Joop and Dalla Vecchia, Claudio and Springel, Volker and Booth, C. M.},
  title = {Feedback and the structure of simulated galaxies at redshift $z=2$},
  journal = {Monthly Notices of the Royal Astronomical Society},
  volume = {409},
  number = {4},
  pages = {1541--1556},
  year = {2010},
  doi = {10.1111/j.1365-2966.2010.17391.x}
}

@article{SchneiderSluse2013,
  author = {Schneider, Peter and Sluse, Dominique},
  title = {Mass-sheet degeneracy, power-law models and external convergence: Impact on the determination of the {Hubble} constant from gravitational lensing},
  journal = {Astronomy \& Astrophysics},
  year = {2013},
  volume = {559},
  pages = {A37},
  doi = {10.1051/0004-6361/201321882},
  eprint = {1306.0901},
  archivePrefix = {arXiv},
  primaryClass = {astro-ph.CO},
  url = {https://www.aanda.org/articles/aa/pdf/2013/11/aa21882-13.pdf}
}

@book{Sersic1968,
  author = {S{\'e}rsic, Jos{\'e} Luis},
  title = {Atlas de Galaxias Australes},
  publisher = {Observatorio Astron{\'o}mico, Universidad Nacional de C{\'o}rdoba},
  address = {C{\'o}rdoba, Argentina},
  year = {1968}
}

@article{SomervilleDave2015,
  author = {Somerville, Rachel S. and Dav{\'e}, Romeel},
  title = {Physical Models of Galaxy Formation in a Cosmological Framework},
  journal = {Annual Review of Astronomy and Astrophysics},
  year = {2015},
  volume = {53},
  pages = {51--113},
  doi = {10.1146/annurev-astro-082812-140951},
  eprint = {1412.2712},
  archivePrefix = {arXiv},
  primaryClass = {astro-ph.GA},
  url = {https://www.annualreviews.org/content/journals/10.1146/annurev-astro-082812-140951}
}

@article{Speagle2020Dynesty,
  author = {Speagle, Joshua S.},
  title = {{dynesty}: a dynamic nested sampling package for estimating Bayesian posteriors and evidences},
  journal = {Monthly Notices of the Royal Astronomical Society},
  volume = {493},
  number = {3},
  pages = {3132--3158},
  year = {2020},
  doi = {10.1093/mnras/staa278},
  archivePrefix = {arXiv},
  eprint = {1904.02180}
}

@article{Springel2001,
  author = {Springel, Volker and White, Simon D. M. and Tormen, Giuseppe and Kauffmann, Guinevere},
  title = {Populating a cluster of galaxies---I. Results at z=0},
  journal = {Monthly Notices of the Royal Astronomical Society},
  volume = {328},
  pages = {726--750},
  year = {2001},
  doi = {10.1046/j.1365-8711.2001.04912.x}
}

@article{Springel2010,
  author = {Springel, Volker},
  title = {{E pur si muove}: Galilean-invariant cosmological hydrodynamical simulations on a moving mesh},
  journal = {Monthly Notices of the Royal Astronomical Society},
  volume = {401},
  pages = {791--851},
  year = {2010},
  doi = {10.1111/j.1365-2966.2009.15715.x}
}

@article{Suyu2006Bayesian,
  author = {Suyu, Sherry H. and Marshall, Philip J. and Hobson, Michael P. and Blandford, Roger D.},
  title = {A Bayesian analysis of regularized source inversions in gravitational lensing},
  journal = {Monthly Notices of the Royal Astronomical Society},
  volume = {371},
  number = {2},
  pages = {983--998},
  year = {2006},
  doi = {10.1111/j.1365-2966.2006.10733.x}
}

@article{Treu2010,
  author = {Treu, Tommaso},
  title = {Strong Lensing by Galaxies},
  journal = {Annual Review of Astronomy and Astrophysics},
  volume = {48},
  pages = {87--125},
  year = {2010},
  doi = {10.1146/annurev-astro-081309-130924}
}

@article{Treu2011SWELLS,
  author = {Treu, Tommaso and Dutton, Aaron A. and Auger, Matthew W. and Marshall, Philip J. and Bolton, Adam S. and Brewer, Brendon J. and Koo, David C. and Koopmans, L{\'e}on V. E.},
  title = {The {SWELLS} survey -- I. A large spectroscopically selected sample of edge-on late-type lens galaxies},
  journal = {Monthly Notices of the Royal Astronomical Society},
  volume = {417},
  pages = {1601--1620},
  year = {2011},
  doi = {10.1111/j.1365-2966.2011.19378.x}
}

@article{WarrenDye2003,
  author = {Warren, Stephen J. and Dye, Simon},
  title = {Semilinear gravitational lens inversion},
  journal = {The Astrophysical Journal},
  volume = {590},
  number = {2},
  pages = {673--682},
  year = {2003},
  doi = {10.1086/375132}
}

@article{Zhao2020Bars,
  author = {Zhao, Dongyao and Du, Min and Ho, Luis C. and Debattista, Victor P. and Shi, Jingjing},
  title = {Barred Galaxies in the {IllustrisTNG} Simulation},
  journal = {The Astrophysical Journal},
  volume = {904},
  number = {2},
  pages = {170},
  year = {2020},
  doi = {10.3847/1538-4357/abbe1b}
}

@article{vanAlbadaSancisi1986,
  author = {{van Albada}, T. S. and Sancisi, R.},
  title = {Dark matter in spiral galaxies},
  journal = {Philosophical Transactions of the Royal Society of London. Series A, Mathematical and Physical Sciences},
  year = {1986},
  volume = {320},
  pages = {447},
  doi = {10.1098/rsta.1986.0128},
  url = {https://doi.org/10.1098/rsta.1986.0128}
}

\bsp
\label{lastpage}
\end{document}